\documentclass[useAMS,usenatbib,usegraphicx]{mn2e}

\usepackage{amssymb}

\usepackage[fleqn]{amsmath}

\long\def\symbolfootnote[#1]#2{\begingroup
\def\thefootnote{\fnsymbol{footnote}}\footnote[#1]{#2}\endgroup}

\usepackage{ifthen}

\newcommand{\dd}{ {\rm d} }

\newcommand{\ddv}[3][1]{
            \ifthenelse{\equal{#1}{1}}
                       {\frac{\dd #2}{\dd #3}}
                       {\frac{\dd^{#1} #2}{\dd #3^{#1}}}}

\newcommand{\pddv}[3][1]{
            \ifthenelse{\equal{#1}{1}}
                       {\frac{\partial #2}{\partial #3}}
                       {\frac{\partial^{#1} #2}{\partial #3^{#1}}}}

\title[``Skinny Milky Way'', says Sagittarius.]
        {``Skinny Milky Way please'', says Sagittarius}
\author[Gibbons, Belokurov \& Evans]
   {S.L.J. Gibbons$^1$\thanks{E-mail:~sljg2,vasily,nwe@ast.cam.ac.uk},
    V. Belokurov$^1$ and N.W. Evans$^1$
  \medskip
  \\$^1$Institute of Astronomy, University of Cambridge, Madingley Road,
       Cambridge, CB3 0HA, UK}
\begin{document}

\date{Accepted 2014 September 22.  Received 2014 September 19; in original form 2014 June 6}

\pagerange{\pageref{firstpage}--\pageref{lastpage}} \pubyear{2014}

\maketitle

\label{firstpage}

\begin{abstract}
Motivated by recent observations of the Sagittarius stream, we devise
a rapid algorithm to generate faithful representations of the centroids of stellar
tidal streams formed in a disruption of a progenitor of an arbitrary
mass in an arbitrary potential. Our method works by releasing swarms
of test particles at the Lagrange points around the satellite and
subsequently evolving them in a combined potential of the host and the
progenitor. We stress that the action of the progenitor's gravity is
crucial to making streams that look almost indistinguishable from the N-body
realizations, as indeed ours do. The method is tested on mock stream
data in three different Milky Way potentials with increasing
complexity, and is shown to deliver unbiased inference on the Galactic
mass distribution out to large radii. When applied to the observations
of the Sagittarius stream, our model gives a natural explanation of
the stream's apocentric distances and the differential orbital
precession. We, therefore, provide a new independent measurement of
the Galactic mass distribution beyond 50 kpc. The Sagittarius stream
model favours a light Milky Way with the mass $4.1 \pm 0.4 \times
10^{11} M_{\odot}$ at 100 kpc, which can be extrapolated to $5.6 \pm
1.2 \times 10^{11} M_{\odot}$ at 200 kpc. Such a low mass for the
Milky Way Galaxy is in good agreement with estimates from the
kinematics of halo stars and from the satellite galaxies (once Leo I
is removed from the sample). It entirely removes the \lq\lq Too Big To
Fail Problem''.
\end{abstract}

\begin{keywords}
Galaxy: halo -- Galaxy: fundamental parameters -- Galaxy: kinematics
and dynamics -- galaxies: dwarf: Sagittarius
\end{keywords}

\section{Introduction}\label{sec:introduction}

Over many years, the uncertainty with which the total mass of the Milky Way is
known has been vexing theorists and observers alike. The spread of allowed
masses covers a large range of possibilities in which both light
($<10^{12}M_\odot$) and heavy ($>2\times10^{12} M_\odot$) Galaxies are permitted
to exist. Attempts to gauge the Galactic matter budget have been made using
a variety of stellar kinematic tracer populations \citep[see e.g.][]{Smith2007,
Xue2008, Bovy2012}. However, these methods suffer from systematics caused by
the lack of reliable tangential velocity and distance measurements. For
non-rotating populations, this is exacerbated by the paucity of fast-moving
and/or distant tracers. For a dataset covering a large variety of sight-lines,
the need for tangential velocities for every star can be alleviated by using an
assumption as to the velocity anisotropy for the population as a whole. The
line of sight velocity contains contributions from both the radial and
tangential velocities as judged from the Galactic centre. Provided the
population is relatively nearby, say at most $\sim 50$ kpc away, then the
velocity anisotropy can be usefully constrained and some of the degeneracy in
mass estimates broken~\citep{Deason2012BD}.

Other attempts at mass measurement have used the motion of the
population of the Galactic dwarf satellites and globular
clusters \citep[e.g][]{Little1987, Zaritsky1989, Wilkinson1999,
Sakamoto2003, Watkins2010}. These have the capability to probe the
mass of the Milky Way to much larger radii. Nonetheless, many Galactic
satellites do not possess reliable proper motions, and in any case the
number of such objects is limited. A further problem is the ambiguous
position of the Leo I dwarf satellite, which has a large line of
sight velocity and Galactocentric distance~\citep{Sohn2013}.  Leo I,
if included, contributes $\sim 30\%$ to \citet{Watkins2010}'s estimate
of the virial mass of the Milky Way. However, there is an element of
circularity in this argument, as once Leo I is assumed to be bound,
the mass of the Milky Way must be large enough to bind it ($\sim
2 \times 10^{12} M_\odot$) !

On smaller scales, the tidal streams emanating from the satellites of
the Milky Way have been suggested as useful constraints on the
mass. This has been done most memorably by \citet{Koposov2012} in
their modelling of the GD-1 stellar stream. Note that the GD-1 stream
is only $\sim 15$ kpc from the Galactic centre, so their analysis
constrains the mass out to modest Galactocentric distances. What is
needed is the modelling of a much more gigantic structure that reaches
out to much greater distances.

Recently, \citet{Belokurov2014} have demonstrated that the trailing arm
\citep[see e.g.][]{SGB-trailing-paper, Sgr3D, Slater2013} of the Sagittarius
(Sgr) stream can be traced out to its apocentre at $\sim100$~kpc. This confirms
the earlier discovery of \citet{Newberg2003} and complements the earlier
detections of the leading debris, whose apocentre lies at $\sim50$~kpc
\citep{field-of-streams}. The vast scale of the Sgr stream has therefore only
recently become apparent. It spans an enormous range of Galactocentric radii,
unparalleled when compared to other known Milky Way streams and substructures.
Thus, the Sgr stream gives us a unique opportunity to make a precision
measurement of our Galaxy's mass out $\sim 100~{\rm kpc}$, far further than
hitherto possible.

However, to carry this out program requires the development of new
modelling techniques. In the past, N-body simulations have been fitted
to Sgr stream data with some success
\citep[e.g.][]{Fellhauer2006,LM2010}, but they suffer from the
drawback that they are extremely time-consuming and so preclude a full
exploration of parameter space. Other attempts at potential inference
using the Sgr stream \citep[e.g.][]{DegWidrow2013,vera2013,ibata2013},
while typically faster than the full-blown N-body computations, have
not tackled the problem of producing realistic looking streams.

The main aim of this paper is to develop an approximate method of quickly
generating streams with realistic centroids by stripping stars at the tidal
radius of a progenitor.  Their evolution in the Galactic gravitational
potential gives us the morphology of the stream, and in particular the
locations of the apocentres of the leading and trailing branches. This enables
us to search through a large class of models to constrain the mass of the Milky
Way.

The structure of the paper is as follows. In Section 2, we summarize the data
on the Sgr Stream. Section 3 introduces a new way to generate streams and
thence to perform inference on the Milky Way potential, which we term modified
Lagrange Cloud Stripping (mLCS). The mLCS algorithm is tested against
simulations of disruption in realistic Milky Way-like potentials in Section 4.
Finally, in Section 5, we apply this method to the observations of the Sgr
Stream to recover an estimate of the mass profile of the Milky Way. Section
6 summarises our results and their implications for conundrums such as the
  ``Too Big To Fail Problem''.

\section{The Data}\label{sec:data}

\begin{table}
\centering
\begin{minipage}{110mm}
\caption{The Sagittarius dwarf at a glance}
\begin{tabular}{llr}
\hline
Property & Value & Ref\\
\hline
Galactic latitude $b$   & $-14\degr.1669$ & [1]\\
Galactic longitude $l$  & $5\degr.5689$   & [1]\\
Heliocentric distance $d_{\sun}$ & 22.0 to 28.4 ${\rm kpc}$& [2]\\
Heliocentric LOS velocity $v_{r\sun}$ & $153\pm2~{\rm km \, s^{-1}}$ & [3]\\
Proper motion $\mu_b$ & $1.97 \pm 0.3 ~{\rm mas ~ yr^{-1}}$ &[4]+[5] \\
Proper motion $\mu_l \cos b$ & $-2.44 \pm 0.3~{\rm mas ~ yr^{-1}}$ &[4]+[5]\\
\hline
{[1] }{\citet{Majewski2003}} & \multicolumn{2}{l}{[2] \citet{Kunder2009}}\\
{[3] }{\citet{Ibata1994}}& \multicolumn{2}{l}{[4] {\citet{Pryor2010} }}\\
{[5] {\citet{Dinescu2005}} }& \multicolumn{2}{l}{~}\\
\end{tabular}
\label{SagTable}
\end{minipage}
\end{table}

The most direct way of inferring the Galactic gravitational potential
is by modeling the paths of test particles orbiting in it. Such
inference, however, suffers from degeneracies if only a small section
of a single orbit is observed
\citep[see e.g.][]{Eyre2009}. In spherical potentials, the rosette pattern of
an orbit can be uniquely described by the size and eccentricity of its
``petals'' -- or equivalently the apocentric and pericentric distances
and the azimuthal precession rate \citep[see
e.g.][]{GalacticDynamics}.

In the Milky Way, of all known stellar tidal streams, only the Sgr stream has
data covering two nearly complete orbital loops, one for the leading tail and
one for the trailing. As \citet{Belokurov2014} show, the apocentre of the
leading tail is firmly placed at $\sim$50 kpc and the trailing debris are
revealed to reach the maximal distance of $\sim$100 kpc from the Galactic
centre. The opening angle, as viewed from the centre of the Galaxy, between the
positions of the respective apocentres is measured to be $\sim 93^{\circ}$.
The apocentric distances of the two tails contain information on the extent of
the progenitor's orbit. Importantly, these also reflect the differences in
energy and angular momentum between the orbit and the debris, which is
essential for the stream modelling.  The differential precession is controlled,
to first order, by the radial mass profile in the Galaxy, but also is a weak
function of the Sgr's orbital eccentricity. Finally, to complete the model, we
require the location of the progenitor in both position and velocity space.
These values will be taken from the literature. In essence, to measure the
total mass distribution in the Galaxy, we aim to reproduce the apocentric
distances together with the differential precession angle $(r_{\rm aL}, r_{\rm
aT}, \Delta \psi)$ of the Sgr stream.

Our best knowledge of the Sgr's current position and velocity is
summarised in Table~\ref{SagTable}.  Where there is disagreement in
the literature as to the value of some of these properties, we have
tried to be as conservative as possible. This is especially evident in
the determination of the heliocentric distance \citep[see][for a
  summary of recent measurements]{Kunder2009}, which we assume lies
somewhere between the extrema of values measured (22.0 -
28.4~kpc). The two best estimates of the progenitor's proper motion
\citep{Dinescu2005,Pryor2010} agree reasonably well with one another,
so we take the mean of their determinations as a best estimate.

We take care in converting from the observable space of
heliocentric distance and proper motions to the true Galactocentric
distance and velocities to avoid introducing any biases.  For the
peculiar motion of the Sun with respect to the Local Standard of Rest
(LSR), we take $(U,V,W) = (11.1,12.24, 7.25)$ ~${\rm km\,s^{-1}}$ from
\citet{Schonrich2010}. For the distance from the Sun to the Galactic
center, we use $R_0 = 8~\rm{kpc}$ and we take the circular
speed~\footnote{Note that we do not tie the circular speed at the
  Solar radius to the fitted halo model as this induces a bias at
  small Galactocentric radii of the order of 10\%.} at $R_0$ to be
$v_c(R_0) = 237~{\rm km~s^{-1}}$.

\section{The Model}\label{sec:model}

\begin{figure*}
\centering
\includegraphics[width=1.0\textwidth]{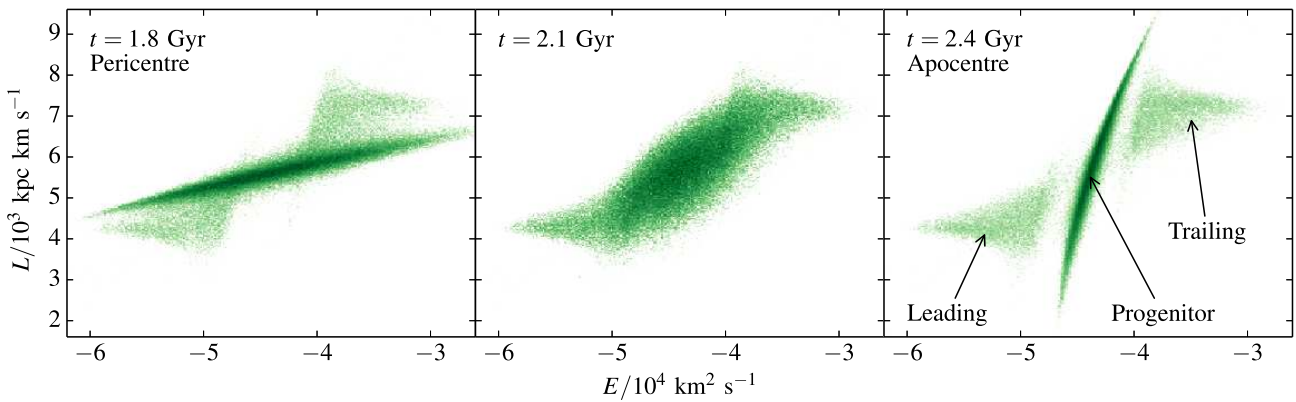}
\caption{\label{fig:EL_intro} The formation of a tidal stream in the
  energy and angular momentum space. Each of the three panels show
  snapshots of an N-body simulation in $E$ and $L$ space at different
  time steps along the progenitor's orbit, namely at pericentre
  (left), between pericentre and apocentre (middle) and at apocentre
  (right). The progenitor forms an elongated pseudo-ellipse which
  rotates (and changes its aspect ratio) as the orbital phase of the
  satellite grows. In this diagram, the unbound debris are seen as two
  fans offset from the $E$ and $L$ of the progenitor's centre of
  mass. The cloud with greater $E$ and $L$ is the trailing arm and
  the fan with lower $E$ and $L$ is the leading arm. Once unbound
  the positions of stars in the ``bow-tie'' are frozen (bar possible
  small changes of the angular momentum in non-axisymmetric
  potentials).}
\end{figure*}

Motions of bodies in a gravitational field are deterministic. Thus, in
principle, given the full velocity and position information for only 2
points on the orbit, initial and final, the underlying potential can
be constrained. This inspires {\it Rewinder} \citep{rewinder}, which
takes advantage of the fact that the initial conditions for stars in
the tidal debris are fully specified if the progenitor is identified
and its orbit is known. Similarly, \citet{hvshalo} show that it might
be possible to constrain the Galactic halo triaxiality based on the
current positions and velocities of hypervelocity stars, assuming
these originate in the vicinity of the central black hole. In reality,
we rarely have accurate measurements of all six phase-space
coordinates across large distances in the Galaxy. Instead, we
typically attempt to leverage the precise knowledge of some of the six
dimensions along a length of an orbit to simultaneously infer the
initial conditions as well as the properties of the force field.

Stellar streams follow their progenitors' orbits approximately.  That
is to say, there exists an offset between the satellite's orbit and
each of the two tidal arms of the stream. Quite simply, the stars in
the debris are launched with initial conditions that are slightly
different from those of the progenitor. If sphericity of the
underlying potential is assumed, then the misalignment between the
stream and the orbit can be modelled simply by using the debris energy
and angular momentum distributions \citep[e.g.][]{Johnston1998}. In
general, this orbit-stream deviation \citep[see e.g.][]{Eyre2010} can
be stipulated in terms of a change in actions and phases between the
progenitor and the tidal debris \citep[e.g.][]{Sanders2013}. As
\citet{Eyre2010} illustrate, stars bound to the progenitor form an
ellipse in action-space, whose flattening and orientation are tied to
the orbital phase of the progenitor (see their Figure 10). As the
progenitor moves along the orbit and the ellipse rotates, the stripped
stars are frozen in action-space in those exact configurations they
were at the time of unbinding. As the disruption continues, the
segments of the rotating ellipse contributed by the stripped stars
start to overlap, forming a bow-tie pattern. This picture clearly
indicates that there is an intricate link between the time of
disruption and the orbital parameters of the stripped stars. However,
the complexity of the debris properties is largely concealed if the
disruption is visualized in frequency space. Here, the leading and the
trailing tail distributions look nearly one-dimensional and can be
modelled as Gaussian \citep[see e.g.][]{Bovy2014, Sanders2014}.

Based on the pioneering ideas of \citet{HelmiWhite}, such stream
models built in action-angle and frequency space have been shown to
work extremely well for cold streams, i.e. those originating from
progenitors with masses up to $10^7-10^8 M_{\odot}$ \citep[for
  discussion, see e.g.][]{Bovy2014}. For more massive systems, the
assumption that the distribution of debris in action space can be
approximated by a Gaussian and that the frequency offset between the
progenitor and the stream is constant throughout the disruption are
likely to break down. Additionally, above $10^8 M_{\odot}$ the gravity
of the progenitor is bound to muddle the elegant predictions based on
the action-angle formalism. According to \citet{no2010}, before the
disruption, the progenitor of the Sgr stream had in excess of
$10^8M_{\odot}$ in stellar mass alone. It is expected that the total
mass of the Sgr dwarf was even higher, quite possibly surpassing
$10^9M_{\odot}$. Therefore, to analyse the properties of the Sgr tidal
tails, we strive to build a stream model that works equally well for
progenitors of any mass.

\subsection{Tidal Stream Mechanics}

\begin{figure*}
\centering
\includegraphics[width=0.495\textwidth]{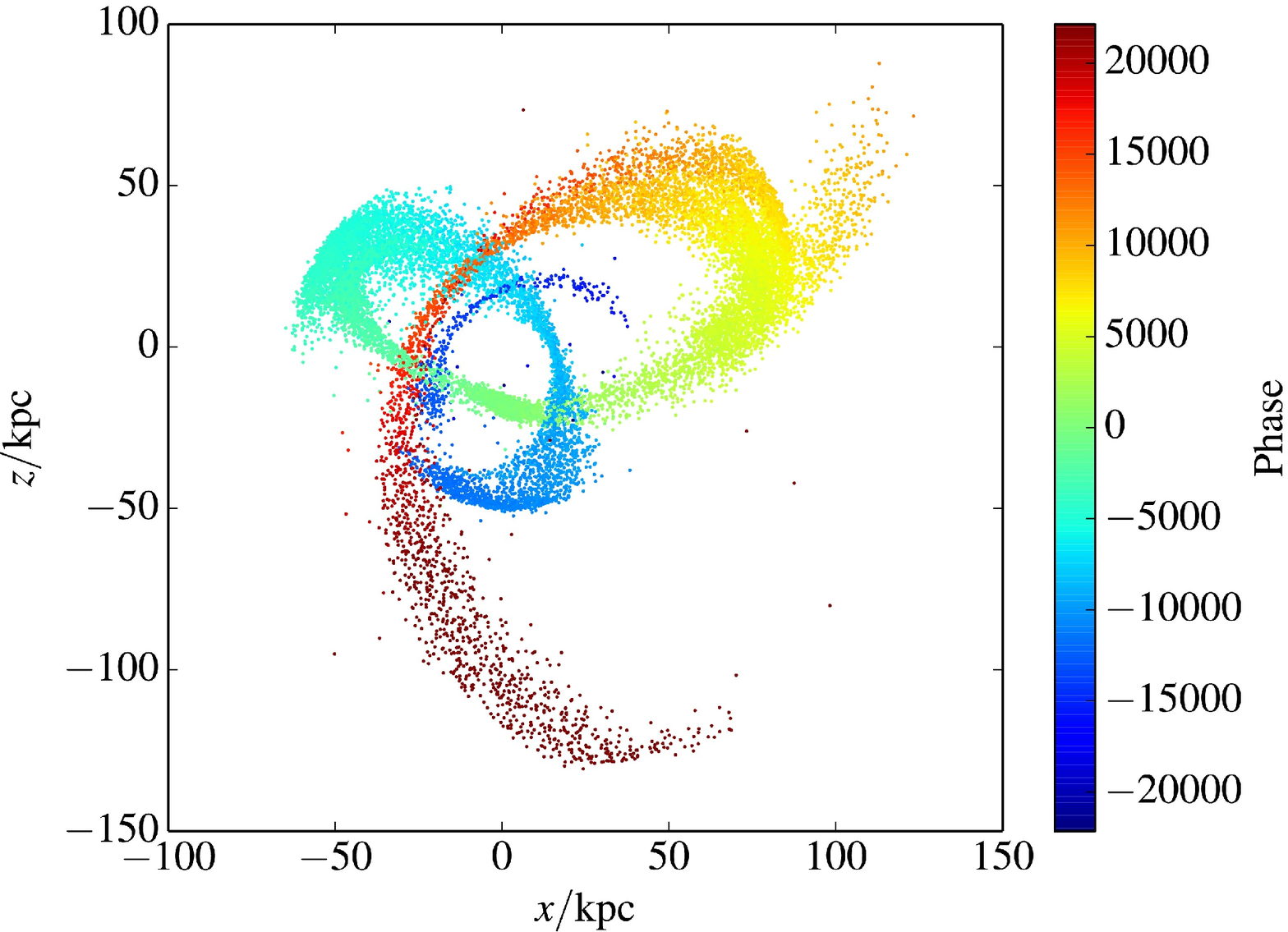}
\includegraphics[width=0.495\textwidth]{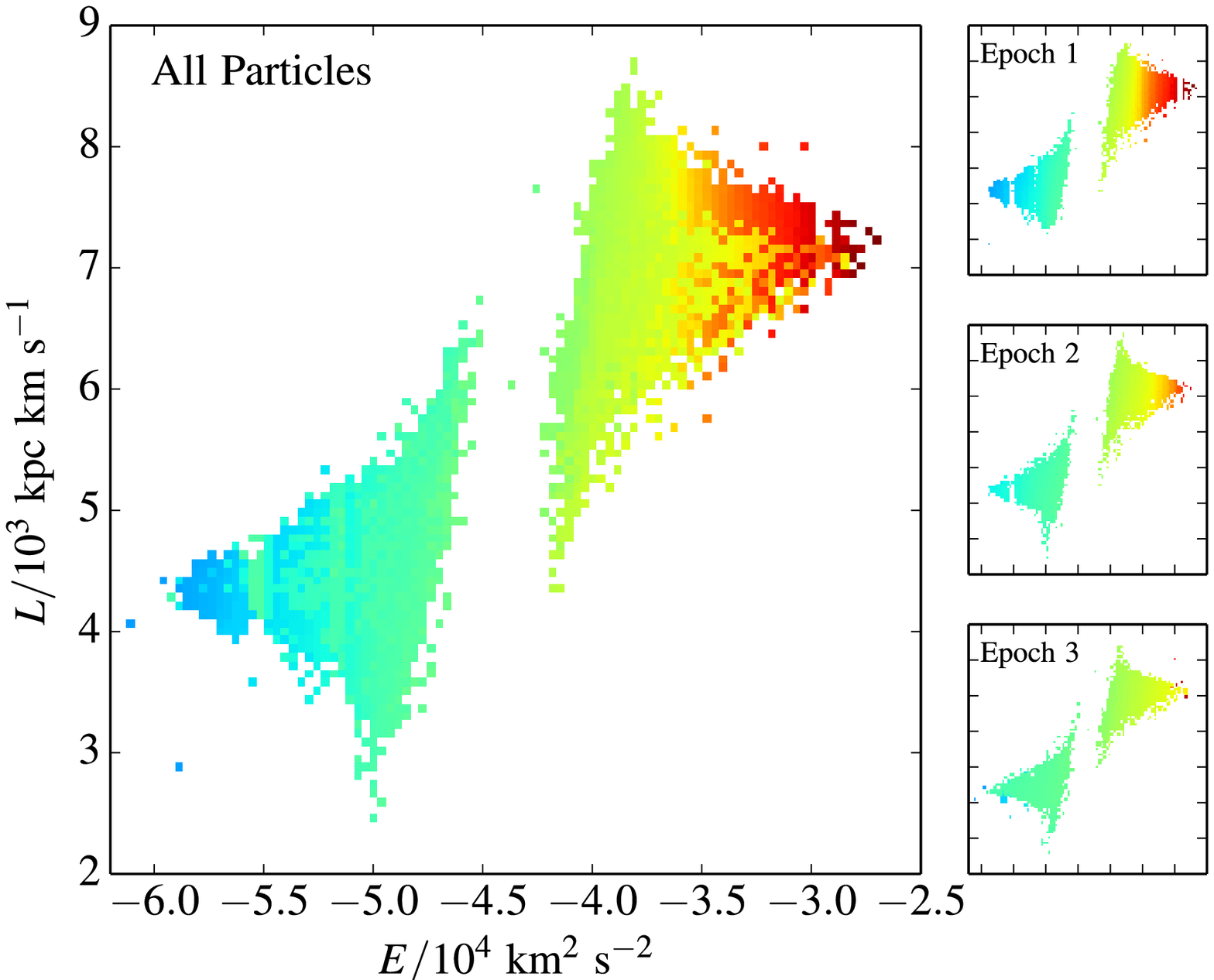}
\caption{\label{fig:phase_colored_dedl} {\it Left:} Distribution of
  the tidal debris produced in the N-body simulation in configuration
  space. The particles are coloured by ``phase'', a metric
  monotonically increasing along the stream. This particular snapshot
  of the simulation corresponds to the fourth pericentric passage,
  some 4 Gyr after the onset of disruption.{\it Right:} Distribution
  of the debris in $E$ and $L$ space. The colouring scheme applied to
  the stream ``bow-tie'' is the same as in the Left panel. The
  particles are binned in $E$ and $L$ space and the median value of
  the phase in each bin is shown. The main panel shows all stream
  particles created up to this time-step. The mini-panels on the right
  show the distributions of particles created at the previous
  stripping epochs. Note that particles created the earliest (Epoch 1)
  have had the most time to spread along the stream, thus creating a
  strict energy sorting. Even though particles from different
  individual epochs overlap in the configuration space, the energy
  sorting persists as evidenced by the main panel.}
\end{figure*}

The objective of this paper is to explore the many-dimensional space
spanned by the parameters controlling the mass profile in the Galaxy,
while simultaneously fitting for the appropriate progenitor
model. Therefore, our method of producing tidal tails must be as fast
as possible while maintaining the necessary degree of likeness when
compared to a ``gold standard'' stream. But what would such a gold
standard be in the case of the Sgr ?

Given the estimated mass of the dwarf prior to disruption and its
orbital period, it seems appropriate to have a ``live'' (as opposed to
a static parameterized density-law) N-body and hydrodynamical model
for both the host and the satellite. The need for the inclusion of not
only the dark matter, but also gas and stars is dictated by the
complexity of the available Sgr stream data. Both leading and trailing
tails show strong metallicity gradients and evidence for star
formation that possibly ceased only a few Gyrs ago. Moreover, each of
the tails seems to be bifurcated into a bright and faint component
\citep{field-of-streams,KoposovSgr}. The very few studies that address
the problem of the disruption of a system with multiple components do
find that it can affect the resulting tidal tails significantly
\citep[see e.g.][]{sgrdisk,taletwicetold}. Similarly, there are many
reasons for the Milky Way model itself to be ``live''. First, this
would naturally account for the dynamical friction effects. Second,
the Sgr disruption could have been going on for so long \citep[see
  e.g.][]{Fellhauer2006} that the Galaxy's mass distribution has
evolved. Finally, not all combinations of the disc and the dark halo
that can be written down in a parameterized form are actually
stable. However, as \citet{Debattista2013} show, in a ``live'' Milky
Way the disc and the central parts of the halo would adjust to each
other's presence to form a long-lived configuration.

As far as we are aware, such an in-depth study of the Sgr disruption
does not yet exist. Instead, the most widely cited recent
analysis~\citep{LM2010} relies on N-body modelling in which the host
potential is parameterized (and fixed), while the single-component
satellite is represented with particles. While not the ideal
specification, this particular simulation does reproduce the largest
number of observables in both the progenitor and the stream to
date. Thus, we attempt to build a rapid modelling algorithm that
reproduces the salient features of tidal streams produced in N-body
simulations such as that of \citet{LM2010}. Our goal, however, is to
speed the stream production by several orders of magnitude.

Before we proceed to assemble the stream model, let us have a glance at
a typical satellite disruption. The progenitor's evolution in the external
gravitational potential is simply understood in the action-angle and frequency
space, but it is equally transparent when looked at in energy and angular
momentum ($E$, $L$) space. Fig.~\ref{fig:EL_intro} shows the results of the
simulation of the disruption of a satellite modelled as a Plummer sphere with
$6.4 \times 10^8 M_{\odot}$ and $a = 0.85$ kpc in a spherical NFW potential
\citep{NFW1996} with the mass $7.5 \times 10^{11} M_{\odot}$ and the
concentration $c = 20$. The satellite is represented with $10^5$ particles and the optimal smoothing length was chosen according to the prescription of \citet{Dehnen2001}. The
orbit has apocentre at 70 kpc and pericentre at 18 kpc and was evolved for
4.31~Gyr ($\sim 3.5$ orbital periods). These choices of parameters for the satellite and its orbit are chosen to mimic those of the Sgr dwarf. The satellite disruption simulation was
carried out using the Gadget-2 code \citep{Springel2005}. This utilises
a tree algorithm to calculate the forces between each of the particles in the
simulation. We have modified the code to implement a static Milky Way
potential by adding an additional force component, dependent on position, at
each force computation. 

\begin{figure*}
\centering
\includegraphics[width=1.0\textwidth]{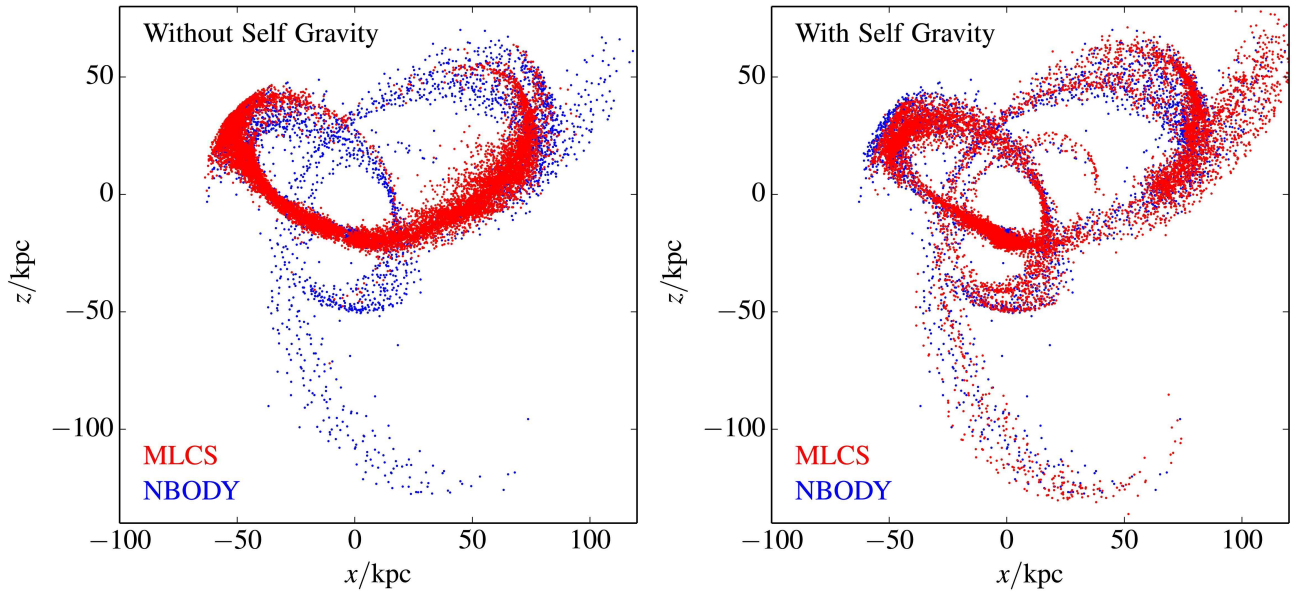}
\caption{\label{fig:gravity_comparison} Comparison between the N-body
  simulation (blue) and the Modified Lagrange Cloud Stripping method
  (red), excluding (left) and including (right) the effects of the
  progenitor's gravity.  The snapshot shown is at 4.31~Gyr after the
  start of the simulation, just after the progenitor's $4^{\rm th}$
  pericentre passage. It is evident that the inclusion of the
  progenitor's gravity is required to correctly model warm streams. If
  the progenitor's gravity is not included, the model stream barely
  reaches half of the length of the N-body representation.}
\end{figure*}

Note that the projection of the progenitor's distribution function onto the ($E$, $L$) space forms a shape closely resembling an ellipse. The inclination of the ellipse can be approximated using the expressions for particle excursions in ($E$, $L$) space presented in Equation 10 of \citet{Yoon2011}. For Roche lobe under-filling satellites and satellites with small Roche lobes, the terms with $\Delta R$ can be neglected, giving a simple expression for the ratio of the amplitudes of the energy and the angular momentum variations:

\begin{equation}
\frac{\Delta E}{\Delta L}\simeq\frac{v_{\rm p}}{R_{\rm p}},
\end{equation}

where $v_{\rm p}$ and $R_{\rm p}$ are the current velocity and the
Galactocentric distance of the progenitor. The subsequent
transformation of the particle distribution happens in the manner
identical to the action-space evolution described earlier. As the
pseudo-ellipse rotates, the stripped particles ``freeze-out" and stop
moving within this space. In Fig.~\ref{fig:EL_intro}, these can be
seen best in the right panel (which presents the situation at the
apocentre crossing) where the debris form two fans, offset from the
energy and angular momentum of the progenitor. The leading tail is
composed of particles with lower $E$ and lower $L$, the trailing tail
particles have higher $E$ and higher $L$ relative to the
progenitor. Curiously, given the shape and the extent of the tidal
debris cloud, it barely overlaps with the progenitor's ellipse at the
pericentre and the apocentre. Therefore, one would conclude that very
little material is stripped at the pericentric crossing, and almost
none at the apocentric.

As Fig.~\ref{fig:EL_intro} illustrates, there is a tight correlation
between the initial conditions of the debris and the orbital phase of
the progenitor at stripping. In other words, after just one radial
period, the stars in the stream will be sorted according to their
energy, with the most distant portions of the tails populated by stars
with the largest differences in energy with respect to the
progenitor. As the satellite completes the subsequent revolutions on
its orbit, the stars in the freshly unbound debris are launched with
the same ($E$, $L$) properties as the previously torn-off stars, but
from locations much closer to the progenitor as compared to the
current positions of the earlier stripped material. Therefore, as the
disruption progresses and more stars are being pumped into the tails,
the strict energy sorting is blurred: at each location along the
stream, the debris posses a mixture of $\Delta E$ and $\Delta
L$. However, depending on the mass of the progenitor, which determines
the extent of the $\Delta E$, $\Delta L$ distributions and its orbital
period, the debris might not have enough time to fully mix.

Fig.~\ref{fig:phase_colored_dedl} shows exactly how much mixing can be
expected for a Sgr-like disruption. The left panel of the Figure shows
the distribution of the debris in the orbital plane, color-coded
according the particle offset from the progenitor as measured along
the stream. The right panel shows the familiar ``bow-tie'' pattern of
debris in ($E$, $L$) space color-coded using the same notation. This
particular snapshot of the simulation corresponds to the fourth
pericentric passage, some 4 Gyr after the onset of disruption. Up to
now, the satellite has experienced 3 bouts of tidal stripping and the
extent of the debris unbound in each round can be gleaned from the
three mini-panels accompanying the right panel of the Figure. These
show the obvious: stars stripped as recently as 1 period ago,
(i.e. Epoch 3) have not had much time to travel far enough along the
stream. On the other hand, the stars stripped in the very beginning of
the disruption (Epoch 1) now cover the entire extent of the tidal
tails. Note however, that notwithstanding the apparent mixing, at each
location along the stream, the superposition of the debris stripped at
different times remains ordered in energy space.

\subsection{Modified Lagrange Cloud Stripping}\label{sec:stream_model}

Naturally, any stream model aiming to closely reproduce the results of
an N-body simulation like the one discussed above must be able to
generate the tidal debris with the correct shape and extent in energy
and angular momentum space. However, this condition is only necessary
but not sufficient, as the stripped stars must also exhibit a certain
amount of correlation between the angle along the stream and ($\Delta
E$, $\Delta L$). These conditions are straightforwardly realised
within the framework outlined by \citet{Varghese2011} and
\citet{Kupper2012} where particles are released from the two Lagrange
points around the progenitor as it moves in the galactic
potential. Here, we utilise a subtly different method as -- rather
than predicting the centroid of the stream directly -- we instead
generate individual stream members. It is from these members that we
calculate the track of the stream and the location of the apocentres
to compare with the observed data. This is very similar to the
methodology used by \citet{Lane2012} in their modelling of the tidal
tails of 47-Tuc, here we will demonstrate the method's utility in
producing streams from progenitors 1000 times more massive. Let us
first explain how the method works in practice and then prove that it
passes the necessary quality checks.

\begin{figure*}
\centering
\includegraphics[width=1.0\textwidth]{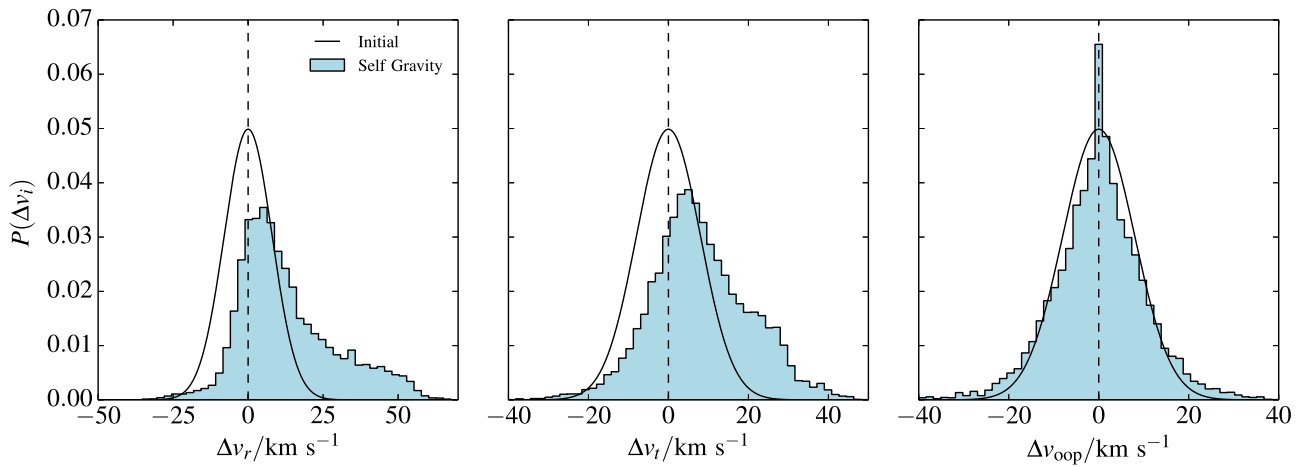}
\caption{\label{fig:strip_vels} Distributions of the velocity
  components of the trailing debris particles at the moment of
  unbinding. The velocity is computed in the frame comoving with the
  progenitor. The panels show the component of velocity along the line
  joining the Galactic centre to the progenitor $\Delta v_r$ (left),
  the component of velocity orthogonal to this in the plane of the
  progenitor's orbit, oriented in the direction of motion of the
  progenitor, $\Delta v_t$ (middle), and the component of velocity out
  of the plane of the progenitor's orbit ,$\Delta v_{\rm oop}$
  (right). Plotted in black is the Gaussian distribution from which
  velocities of stream particles are initially drawn. The velocity
  distribution of stream particles the last time they are at one tidal
  radii from the progenitor is shown by the blue histogram. Note that
  under the action of the satellite's gravity, the distributions of
  velocities become skewed towards higher $\Delta v_t$ and $\Delta
  v_r$, whilst the component of velocity out of the progenitor's
  orbital plane is unaffected.}
\end{figure*}

We start by taking the current position of the progenitor and
integrating it back in time in the assumed galactic potential, $\Phi$,
to its position at a time $t = -t_{\rm back}$ from the present
day. Stream particles are then produced at equal time steps $\delta t$
along the orbit of the progenitor from $t=-t_{\rm back}$ to the
present day. These are launched from two locations on the line joining
the progenitor to the galactic centre at radial offsets $\Delta r =
\pm r_t$, where $r_t$ is the instantaneous tidal radius of the
progenitor defined by:

\begin{equation}
r_t = \left( \frac{G M_{\rm sat}}{\Omega^2 - \frac{\dd^2 \Phi}{\dd r^2}} \right)^{1/3}
\end{equation}

Here $M_{\rm sat}$ is the mass of the progenitor and $\Omega$ is the
instantaneous orbital angular velocity of the progenitor. The
particles released closer to the Galactic centre ($\Delta r < 0$) will
form the leading arm of the stream and the particles set free further
away from the Galactic centre ($\Delta r > 0$) will form the trailing
arm. The velocity of these particles are drawn from a Gaussian
distribution centered upon the velocity of the progenitor with a
dispersion $\sigma$ in each direction. This dispersion is chosen to be
representative of the internal velocity distribution of the stars
within the outer parts of the progenitor. The clouds of stream
particles released at Lagrange points are then evolved within the
combined potential of the host galaxy and the progenitor until the
present day ($t=0$). In our experiments, the progenitor's potential is
treated as a Plummer sphere of fixed mass $M_{\rm sat}$ and scale
radius $a_{\rm sat}$ which moves along the orbit of the progenitor.

Each of the created particles are entirely independent of one other (since the
progenitor's orbit is determined only by the galactic potential) therefore will
scale linearly with the number of particles produced. Thus each particle can be
evolve separately using an adaptive stepping algorithm, naturally accounting
for the differences in dynamical times between the particles which are
recaptured by the progenitor and those which quickly escape. Whilst this
would imply that the mLCS algorithm is trivially parallelizable, we opt not to,
instead using a parallelizable MCMC sampler when exploring parameter space.

Our model requires 3 hidden parameters to describe the structure of
the progenitor: its mass $M_{\rm sat}$, an internal scale length
$a_{\rm sat}$, and the velocity dispersion in the outer parts of the
progenitor $\sigma$. Note that, while in principle one should be able
to calculate the third from any two of these for a self-gravitating
body, we opt to keep all three independent.  This is due to the fact
that the progenitor of the Sgr Stream was presumably embedded in a
dark matter halo as it was accreted, and this will affect the velocity
dispersion of the satellite's stars especially in the outer regions.
The degree of this embeddedness is unknown and therefore to encode our
ignorance we leave all three parameters within the model to be
marginalised over. Note that, additionally, \citet{Kupper2012} require
a special treatment of the tangential component of stream velocity:
having experimented with values between the orbital and the angular
velocity of the progenitor, the settle on the latter. We find this
distinction makes no perceptible difference to the model streams that
we produce and therefore we stipulate, for simplicity, that the debris
mean velocity for both leading and trailing tails is that of the
progenitor.

\begin{figure*}
\centering
\includegraphics[width=0.88\textwidth]{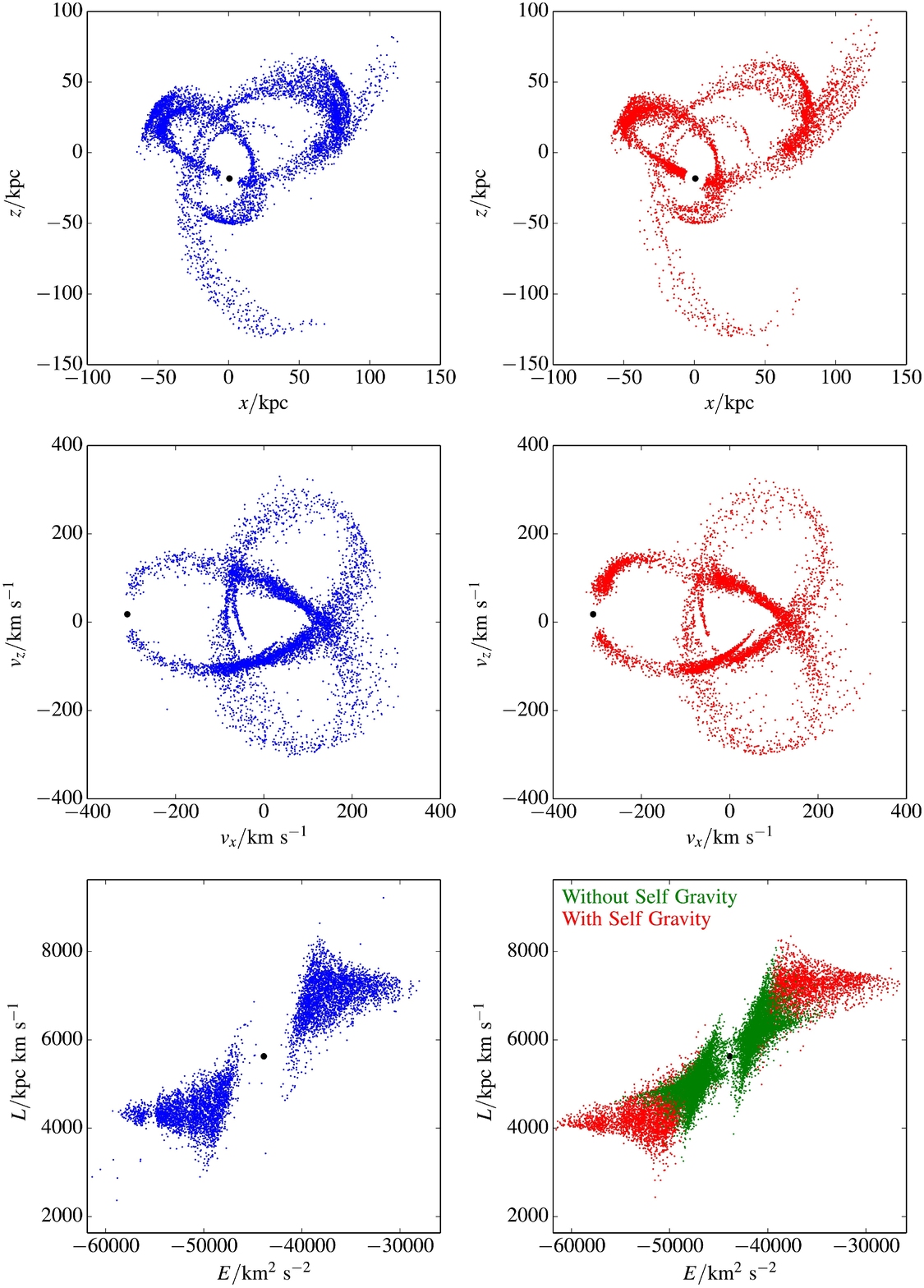}
\caption{\label{fig:NBODY_model_comparison} Comparison between an
  N-body stream (left) and the Modified Lagrange Cloud Stripping model
  (right) in configuration space (upper), velocity space (middle) and
  ($E,L$) space (lower). The location of the progenitor is marked with
  the black filled circle. For ease of comparison we have plotted a
  random selection of the N-body particles so that the same number
  (6500) are plotted for both cases. The bottom right panel also shows
  over-plotted with green points the energy and angular momentum
  distribution of the debris in absence of the progenitor's
  gravity. It is evident that the progenitor's gravity is required to
  assign the correct orbital properties to the debris particles }
\end{figure*}

\begin{figure*}
\centering
\includegraphics[width=0.9\textwidth]{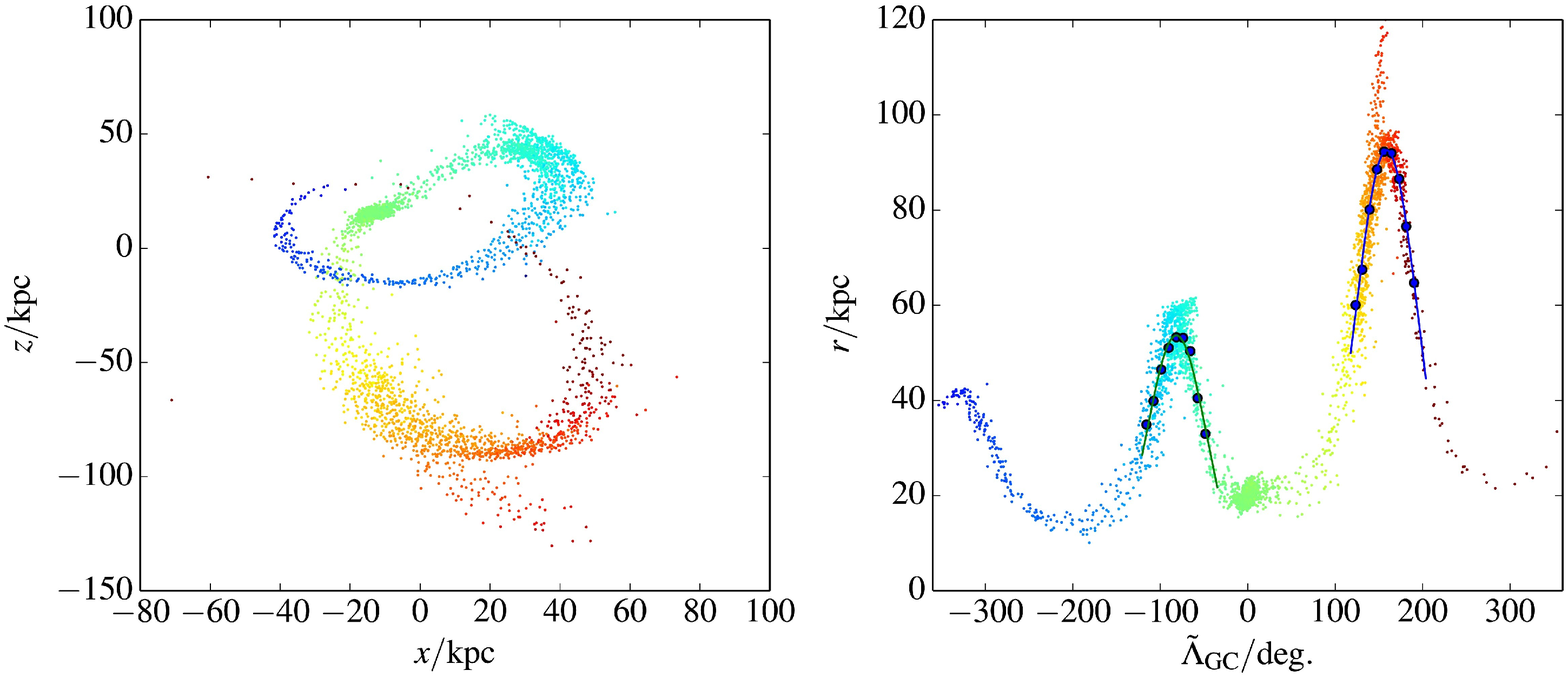}
\caption{\label{fig:unwrap} Illustration of the stream unwrapping
  process and the extraction of the observables. {\it Left:} This
  shows the model stream as in Fig.~\ref{fig:NBODY_model_comparison}
  coloured according to the ``phase'' of the debris. The ``phase'' metric
  increases approximately monotonically along the stream. {\it Right:} This
  shows the same stream unwrapped as a function of angle (on the celestial
  sphere) along the stream, measured from the galactic centre
  $\tilde{\Lambda}_{\rm GC}$ with the progenitor located at
  $\tilde{\Lambda}_{\rm GC} = 0 $. This angle is as defined in
  \citet{Belokurov2014}. The points indicate the centres of the bins used to
  locate the apocentres and the lines show the fitted Gaussians. This allows to
  measure the apocentric distances of the mock stream $r_{\rm aL} = 53.6~{\rm
  kpc}$ and $r_{\rm aT} = 92.6~{\rm kpc}$ as well as the precession angle
  $\Delta \psi = 119^{\circ}.5$
}
\end{figure*}

One crucial difference between our stream model and the previous
implementations based on Modified Lagrange Cloud Stripping is the
inclusion of the progenitor's
gravity. Fig.~\ref{fig:gravity_comparison} compares a stream produced
in N-body simulation (blue points) and two different runs of our model
(red points), without (left) and with (right) the progenitor's
gravity. Without the effects of the satellite's gravity, the tidal
tails seem to possess approximately the correct shape, but drastically
reduced length. This paper is concerned with the modelling of the
positions of the Sgr leading and trailing apocentres. However, having
such stubby tails means that the apocentres would be barely reached by
the stream particles, therefore biasing the model prediction. The
right panel of Fig.~\ref{fig:gravity_comparison} and
Fig.~\ref{fig:NBODY_model_comparison} demonstrates that the inclusion
of the satellite's potential produces a tidal stream almost
indistinguishable from the N-body realization in configuration and
velocity spaces as well as reproducing the overall structure of the
stream in integral of motion space (albeit with a distribution which
is slightly more extended in energy space than the N-body).  The
slight differences appear mostly in the density of stream particles,
which we do not expect to be able to reproduce due to the assumption
of uniform stripping along the orbit. However, as we shall see in
section~\ref{sec:testing}, these minute deviations do not introduce
any significant biases when inferring the galaxy's mass distribution.

Yet, there is one important distinction: the Gadget-2 run took
17.6~CPU~hours (i.e. 2.2 hours wall time on 8 cores), producing 31765 particles
which were unbound from the progenitor and became members of the stream; on
average, this is 2 CPU-seconds per stream particle.  Our Modified Lagrange
Cloud Stripping model only takes 16~CPU~{\it seconds} to run on a single core
of the same server used for the N-body. Here, 12660 particles were created, of
which 9022 ended as stream members. This is
0.002~CPU~seconds per stream particle.  Hence our model shows itself to be
3 orders of magnitude faster than an N-Body to produce the same number of
unbound stream particles. This speed increase is primarily due to the fact
that we do not attempt to resolve the progenitor with live particles, hence
both largely reducing the number of particles that need to be followed, as
well as cutting down the expense of the inter-particle force computations. In
both cases the code was run on a dual socket server using quad core Intel
Xeon X5460 processors, the GADGET-2 run was parallelized to use all 8 cores
whilst the mLCS code ran on only one. Note that the timings presented are for
the production of a \textit{single} stream and not for a parameter search.
Finally it should also be noted that the speed up of this technique over the
N-body run will be dependent on the size, mass and orbit of the progenitor;
the timings presented here are for a progenitor that is typical of a Sgr-like
object, but could differ e.g. for the case of a much larger plummer sphere
the majority of the particles would become unbound almost immediately.

In the next section, we take a closer look at the effects of the
progenitor's gravity which appears to increase the numbers of
particles escaping onto orbits with greater $\Delta E$ and $\Delta L$
as compared to the progenitor itself.

\subsection{Action of the Progenitor's Gravity}

Even though the method laid down in the previous Section appears to be
an over-simplified brute-force imitation of the tidal disruption
process, something curious happens to the clouds of particles released
at the inner and the outer Lagrange points in the presence of the
progenitor's gravity. Fig.~\ref{fig:strip_vels} compares the Gaussian
distributions, from which the particle velocities were drawn, with the
actual velocity distributions at the time they became unbound
from the satellite. The exact unbinding time is not easy to calculate,
therefore instead we use the time of the last tidal radius
crossing. Fig.~\ref{fig:strip_vels} shows the distributions of the three velocity
components of the trailing tail particles. The chosen projections are:
the component of the velocity along the radial vector $v_r$; the
component perpendicular to $v_r$ in the plane of the orbit, pointing
in the direction of motion of the progenitor, $v_t$; and, finally, the
component pointing out of the orbital plane $v_{\rm oop}$. As
Fig.~\ref{fig:strip_vels} illustrates, the leaving population has a
prominent excess in $v_r$ and $v_t$.

With this in mind, it is now easy to explain the discrepancy
identified in the previous Section, namely the production of very
short tidal tails in the absence of the progenitor's gravity. The
bottom right panel of Fig.~\ref{fig:NBODY_model_comparison} gives
the $E$ and $L$ distribution of particles produced both with (red) and
without (green) the satellite's gravity. It is immediately obvious
that in the case without gravity, stream particles are not launched
with sufficiently high energies and angular momenta relative to the
progenitor. Thus, the effect of the satellite's self-gravity is to
modify the crude Gaussian approximation of the velocity distribution
at the time of stripping to a more realistic one. The initial speeds
of the particles drawn form the Gaussian are too low, therefore many
of these are accreted by the progenitor. These recaptured particles
are then evolved internally within the progenitor until the conditions
are favorable for their release. The released particles have an excess
in the radial velocity component as they must be traveling away from
the satellite. The excess of angular momentum is attained by higher
tangential velocities relative to the satellite.

The action of progenitor's gravity explains why, unlike \citet{Kupper2012},
we do not need to tweak the tangential velocity of the released particles
. Our stream production works effectively as a
constrained N-body simulation where we follow the orbital evolution of
large number of massless tracer particles. However, importantly, these
are not necessarily stripped at the time at which they are produced
within the model. Thus, the absolute distribution of velocities given
to stream particles will be erased due to some amount of evolution
within the progenitor's potential. Bear in mind that some accreted
particles are never released. We have estimated that the fraction of
these stuck tracers is mostly of order of 25\% and never above 50\%,
thus the method can be potentially sped up, but only up to a factor of
2.

The obvious advantage of this stream model is that it should work
equally well and equally fast in any host potential, including a live
potential, generated, for example, as part of the cosmological zoom-in
simulation. However, it should be noted that due to the release of the
stream particles at equal times along the progenitor's orbit, our
method will not be able to reproduce the density of stars along the
stream exactly. Despite this, stream particles are located where they
should be in both the configuration and the velocity space.

\subsection{Extraction of Apocentres}\label{sec:extract_apos}

We now devise a method to extract the locations of the stream
apocentres from the generated particle distributions.  The model
streams generally have multiple wraps and thus overlap in space. We
cannot therefore simply use a simple angular coordinate in the orbital
plane to parameterise the location of a particle along the
stream. Instead, we construct a method of explicitly unwrapping the
stream by defining a coordinate which increases monotonically along
the stream.

To do this, we introduce the phase $(\chi)$ of each particle, which is
defined as follows:

\begin{equation}
\chi = \sum  \sqrt{x^2+y^2+z^2} - \sqrt{x_{\rm p}^2 + y_{\rm p}^2 + z_{\rm p}^2}.
\end{equation}

Here, the sum is taken over equal time steps for each stream particle,
from the moment that it is generated in the model up until the epoch
of observation. The subscript p's denote the progenitor.  This
definition works as the particles in the leading (trailing) arm tend
to be at positions which have lower (higher) Galactocentric radii than
the progenitor. Whilst this doesn't provide an entirely monotonically
increasing variable, it does work well enough for our purposes to
unwrap the stream particles and to identify how each segment of the
stream joins together. The results of applying this method are
demonstrated in Figs~\ref{fig:phase_colored_dedl}
and~\ref{fig:unwrap}.

With the model stream unwrapped, we can measure the centroids of the
apocentres along each of the dimensions of the phase-space.  This is
done in a manner which is as close as possible to the method used in
analysing the observational data~\citep{Belokurov2014}.  In general,
more than one apocentre of the stream is produced with this method
(and would be produced in reality as the stream lengthens with
time). The data presented in \citet{Belokurov2014} are almost
certainly from the first wrap of the stream, and thus we consider only
the closest apocentre to the progenitor for both the leading and
trailing arms.  The trailing arm can show a feature in which the
debris stripped from the most recent orbital passage lies on top of
the measured apocentre of interest and must therefore be removed from
the apocentre detection (see the right hand panel of
Fig.~\ref{fig:unwrap}). This can be done by removing any particles
stripped between the last apocentric passage of the progenitor and the
current epoch, where we define the stripping time to be the last time
that the stream particle was within $r_{\rm t}$ of the
progenitor. Next, we bin the stream particles as a function of angle
along the stream to produce an estimate of the stream's centroid. With
this estimate, we then find a first approximation to the positions of
the apocentres as the bins showing a local maximum closest to the
location of the progenitor.  We then take the stream particles within
$\pm 40^{\circ}$ of the initial guess, and fit a Gaussian to the
binned particles in this range\footnote{Whilst any strongly peaked
  function will work adequately here, the choice of a Gaussian is to
  follow \citet{Belokurov2014} as closely as possible in extracting
  the observables, therefore circumventing any additional biases
  introduced by the choice of functional form.}. An example of this
binning and the extraction is shown in the right hand panel of
Fig.~\ref{fig:unwrap}.

\subsection{Galactic Potential Model}\label{sec:galactic_model}

\def\rs{{r_{\rm s}}}

The final ingredient required to carry out the inference is the model
for the Galactic gravitational potential. The HI rotation curve in the
inner Milky Way is flat to a good approximation. Once the HI gas gives
out beyond $\sim 20$ kpc \citep[e.g.][]{Sofue2009}, the rotation curve
may continue flat or may decline. A simple yet flexible family of
models which encompasses such behaviour is:

\begin{equation}
v_{\rm circ}^2 = \frac{v_0^2 \rs^\alpha}{\left(\rs^2 +
  r^2\right)^{\alpha/2}}.
\label{eq:circ}
\end{equation}

The implied Galactic density is

\begin{equation}
\rho(r) = \frac{v_0^2\rs^\alpha}{4\pi G r^2}{\rs^2 +
  (1-\alpha)r^2\over (\rs^2 +r^2)^{1 + \alpha/2}}.
\end{equation}

Naturally, given the sphericity of these models, the enclosed mass as
a function of radius can be found directly from $v_{\rm circ}$:

\begin{equation}
M(r) = \frac{r v_{\rm circ}^2}{G} = \frac{r~v_0^2 \rs^\alpha}{G\left(\rs^2 +
  r^2\right)^{\alpha/2}}.
\end{equation}

\begin{figure}
\centering
\includegraphics[width=0.5\textwidth]{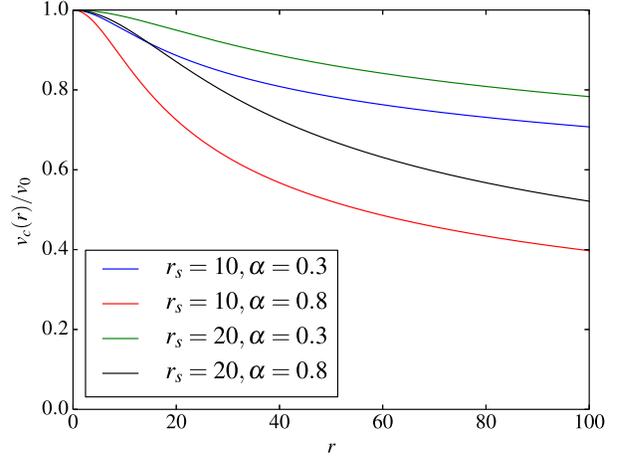}
\caption{\label{fig:rs-alpha} Selection of rotation curves generated
  by the truncated, flat rotation curve (TF) model.  As $\alpha
  \rightarrow 0$, this corresponds to the everywhere flat rotation
  curve of an isothermal sphere.  As $\alpha \rightarrow 1$ and
  $\rs \rightarrow 0$, the model tends towards the Keplerian
  point-mass limit of $v_{\rm circ} \propto r^{-1/2}$.}
\end{figure}

We shall refer to these models as the truncated, flat rotation curve
family (TF). The circular velocity curve (\ref{eq:circ}) is flat with
amplitude $v_0$ in the inner parts and tends to a power-law with slope
$-\alpha/2$ in the outer parts with the transition scale given by
$\rs$. A plot showing the variety of possible rotation curves is shown
in Fig.~\ref{fig:rs-alpha}.  When $\alpha =0$, this is the singular
isothermal sphere.  When $\alpha =1$, this is the model first
introduced by \citet{LinLyndenBell1982} to study the orbits of the
Magellanic Clouds and subsequently used by \citet{Wilkinson1999} in
their measurement of the mass of the Milky Way. As $0 \leq \alpha \leq 1$, the outer rotation curve spans
the physical range from flat to Keplerian. Note, that our model
represents the combined contribution of the disc and the dark halo to
the rotation curve. On the plus side, it is extremely concise, but it
suffers from the obvious drawbacks: it is spherically symmetric and
does not contain a mass component mimicking the central over-density
due to the Galactic bulge/bar. Nonetheless, surprisingly, as we
demonstrate in Section~\ref{sec:testing}, this simple model does not
suffer any significant biases when applied to analyze the behaviour of
streams produced in significantly more complex potentials.

\begin{figure*}
\centering
\includegraphics[width=0.99\textwidth]{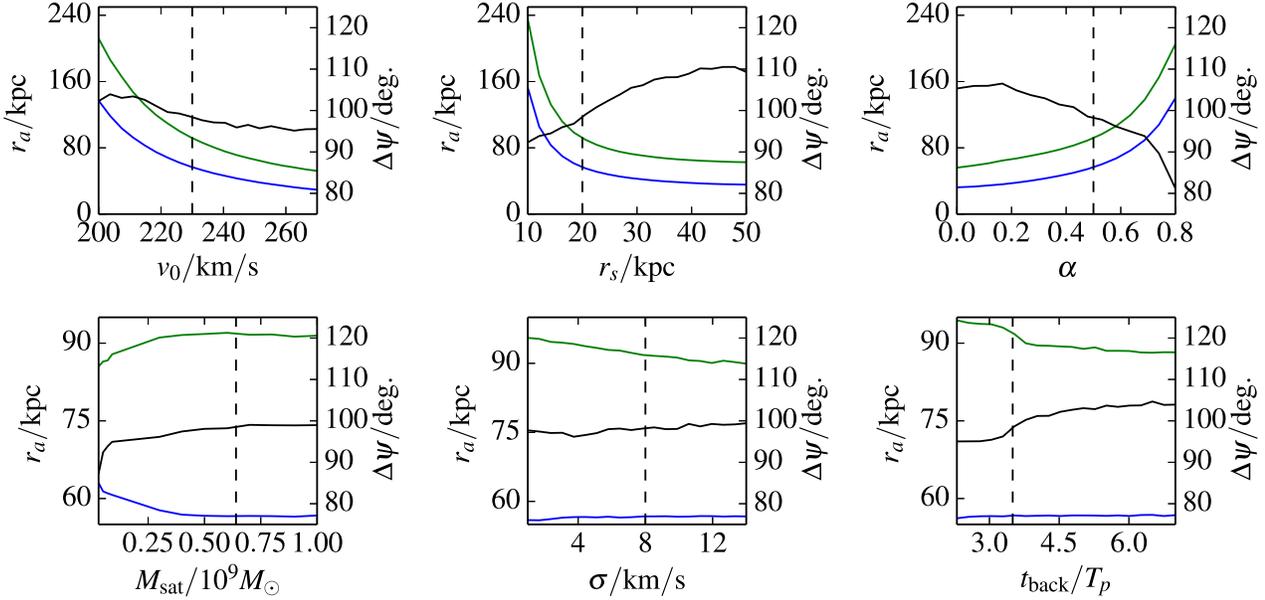}
\caption{\label{fig:model_deps} Dependence of each of the observables
  on one of the model parameters whilst keeping the rest fixed. The
  blue (green) line shows the apocentric distance of the leading
  (trailing) tail $r_{\rm aL}$ ($r_{aT}$), while the black line
  displays the evolution of the value of differential orbital
  precession $\Delta \psi$.  In each panel, vertical dashed line shows
  the value that each of the parameters is fixed to when others are
  varied. Note that the observables are mostly sensitive to changes in
  the potential model. For example, decreasing $v_0$, $\rs$ and
  increasing $\alpha$ all lead to a lighter Milky Way, which helps to
  bring the differential precession down and move the two tidal tails
  apart. Additionally, smaller and subtler evolution of the
  observables is displayed when the nuisance parameters such as
  progenitor's mass and look-back time are varied.}
\end{figure*}

It is illuminating to enquire how sensitive the predicted observables
are to each of the model parameters. We take a progenitor with a fixed
position and velocity today and vary each of the parameters in turn
whilst keeping the others fixed.  The results are shown in
Fig.~\ref{fig:model_deps}.  As the rotation curve becomes more
steeply decreasing in the outer parts ($\rs$ decreases and $\alpha$
increases), we see that the precession angle $\Delta \psi$
increases. This is the same qualitative behaviour one would expect by
treating the stream as an orbit \citep{Belokurov2014}. Fig.~\ref{fig:model_deps} also
gives a clear prediction as to how the inter-arm apocentric distance
difference behaves: lighter Milky Ways, either due to a smaller
circular velocity normalization or due to a steeper fall-off in
density, produce tails that differ the most in their apocentric distances.

Additionally, the difference in apocentric distances decreases as the
stream is allowed to evolve for a longer time. This can be understood
on noting that the stars produced in each of the arms of the stream
have a significant spread in energy (see
e.g. Fig.~\ref{fig:NBODY_model_comparison}).  The stars with the
greatest difference in energy also have the greatest difference in
orbital period as compared to the progenitor. Thus, these end up
furthest from the progenitor along the stream and form stream
apocentres first. As the stream is allowed to evolve for a longer
time, particles with a smaller difference in energy, and thus smaller
difference in apocentric radii, reach apocentre thus causing the
observed effect.  A similar effect can be seen when looking at the
dependence on the observables on the mass of the progenitor $M_{\rm
  sat}$. Here, we see an increase in difference of apocentric
distances on the mass of the progenitor. This is caused by the greater
difference in energy between the two arms of the stream.

\begin{figure*}
\centering
\includegraphics[width=0.33\textwidth]{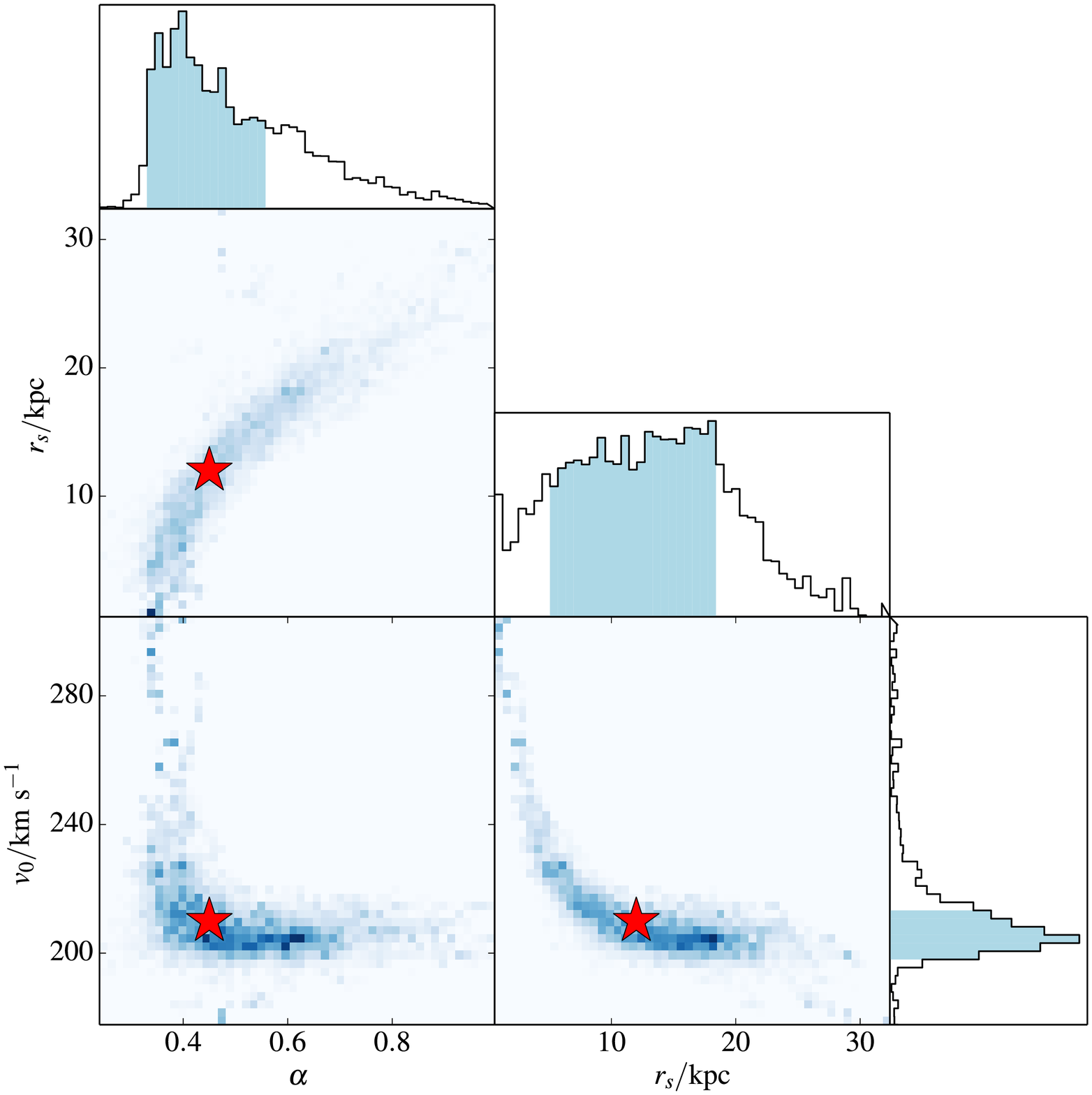}
\includegraphics[width=0.33\textwidth]{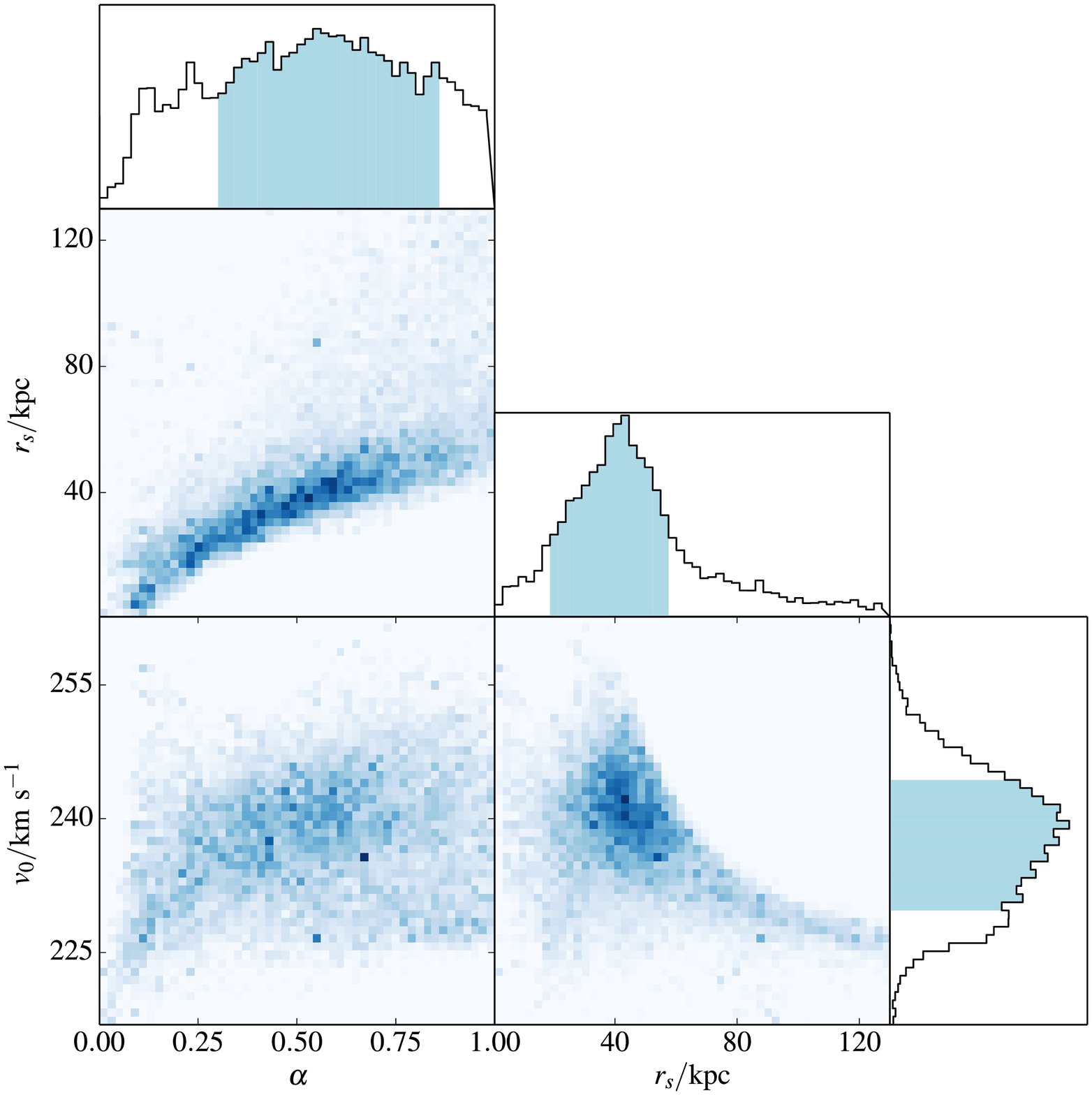}
\includegraphics[width=0.33\textwidth]{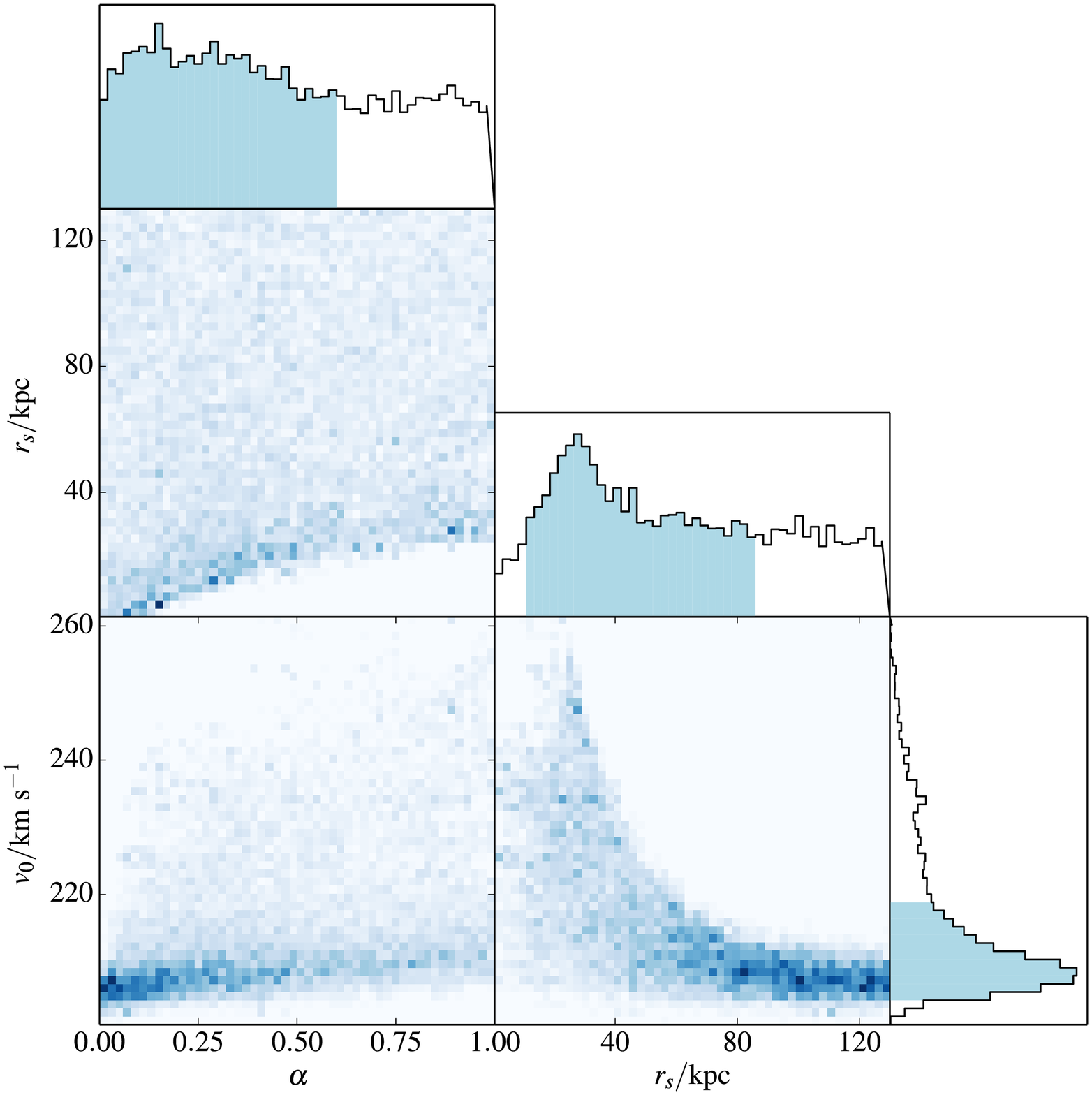}
\caption{\label{fig:potential_recovery} Testing the model on mock
  stream data. Each of the three groups of panels shows the posterior
  probability distributions for $\rs, \alpha$ and $v_0$ when fitting
  the TF model to the stream observables from the test N-body
  simulations. {\it Left Group:} The mock Galaxy is the spherical TF
  model. While there are obvious degeneracies between the model
  parameters, the marginalized 1d posterior distributions peak in the
  close proximity of the input ``true'' values for $\alpha$ and $v_0$
  marked by red stars. The marginalised posterior for $r_s$ is much
  broader, nonetheless the input value is well contained within the
  68\% credible region.  {\it Middle Group:} The mock Galaxy is a
  three-component model with spherical DM halo. Degeneracies similar
  to the earlier case are present. However, it is not straightforward
  to interpret what the peaks in the marginalized posterior
  distributions symbolize.  {\it Right Group:} The mock Galaxy is that
  of \citet{LM2010}. All parameters have long tails in the posterior
  distributions and evidence for strong degeneracies. However, despite
  the obvious model mismatch the cumulative mass profiles are closely
  reproduced (see Fig.~\ref{fig:mass_recovery})}
\end{figure*}

\subsection{Constructing the Mass Estimate}\label{sec:massest}

With a model in place to predict the apocentric properties of a stream
in a given potential, it is now viable to use the observations to
infer the parameters controlling the Galactic mass distribution.  This
is done by comparing the predictions of the model to the observed
values with the following likelihood function:

\begin{equation}
\mathcal{L} = \prod_i \frac{1}{\sqrt{2\pi \sigma_i^2}} \exp
\left[\frac{-(x_{{\rm model,}i}-x_{{\rm data,}i})^2}{2\sigma_i^2}\right]
\end{equation}

where $x_i$ is each of the observables and $\sigma_i$ is the error
associated with it.  The product is taken over each of the three
observables, namely the apocentric distances of the leading and
trailing arms and the precession angle, $(r_{\rm aL}, r_{\rm aT},
\Delta \psi)$

In practice, the model streams are produced by first integrating the
Sgr dwarf orbit back from the location in the phase-space which is
allowed by the current observations. While all 6 coordinates of the
Sgr dwarf have been measured to date, some (for example, the proper
motion) carry more uncertainty than the others (for example, location
on the sky). This is reflected in Table~1 as well as Table~2 which
summarizes the priors used in the modeling. In a nutshell, we are
constraining the total of 3 gravitational potential parameters while
marginalizing over 10 nuisance parameters in total: 6 controlling the
progenitor's orbital initial conditions, 3 controlling its
structural properties, as well as the look-back time

The parameter space is sampled with the parallel Markov Chain Monte Carlo code
{\tt emcee} \citep{emcee} to build a set of samples from the posterior
distributions. From these, it is a simple task to convert into the
estimated mass at a given radius, $M(r)$. We take the joint samples of the
potential parameters and calculate the mass implied by each at the radius of
interest. By using a large number $(\approx 10^5)$ of such samples, we build up
the implied posterior on the mass enclosed at a given radius and thus can
compute the most likely value and its error with ease.

\section{Tests on Mock Data}\label{sec:testing}

To gain confidence in any inferences obtained from our model, we test
it for biases using simulations of disruption within a realistic
Milky-Way like potential.

\subsection{Description of Simulations}\label{sec:sim_description}

The test simulations are performed in three different static Galactic
potentials. Our first mock Galaxy is the same as the model used to
measure the mass distribution, i.e. the truncated, flat rotation curve
family introduced in Section~\ref{sec:galactic_model}. The second case
is a standard three component model of the Milky Way's potential with
bulge, disc and dark matter halo. Here, the central parts of the
potential are flattened due to the disc contribution, but the dark
halo and, therefore, the outer Galaxy is spherical.  Finally, the
third case is the potential proposed by \citet{LM2010} with a
triaxial dark halo.

In each case, the disrupting progenitor is a Plummer sphere of mass
$M_{\rm sat} = 6.4\times 10^8 M_\odot$ and scale-length $a_{\rm sat} =
850$ pc. This yields a satellite with the internal velocity dispersion
of $\sim 30$ kms$^{-1}$ in line with observational constraints of the
Sgr dwarf. The progenitor was placed on a polar orbit with an
apocentre of 75~kpc and a pericentre of 17~kpc and evolved for
$\approx 2.5$ orbital periods ($\sim$3~Gyr) using the Gadget-2 code
\citep{Springel2005}. 

Using the mock N-body observations, it is possible to test the stream
model for any possible biases present in the mass recovery.  Of
course, the simulation data provide perfect knowledge of the current
phase space location of the progenitor. When we apply the model to
the real data, uncertainties in the satellite's coordinates will be
marginalised over.

\begin{figure*}
\centering
\includegraphics[width=0.33\textwidth]{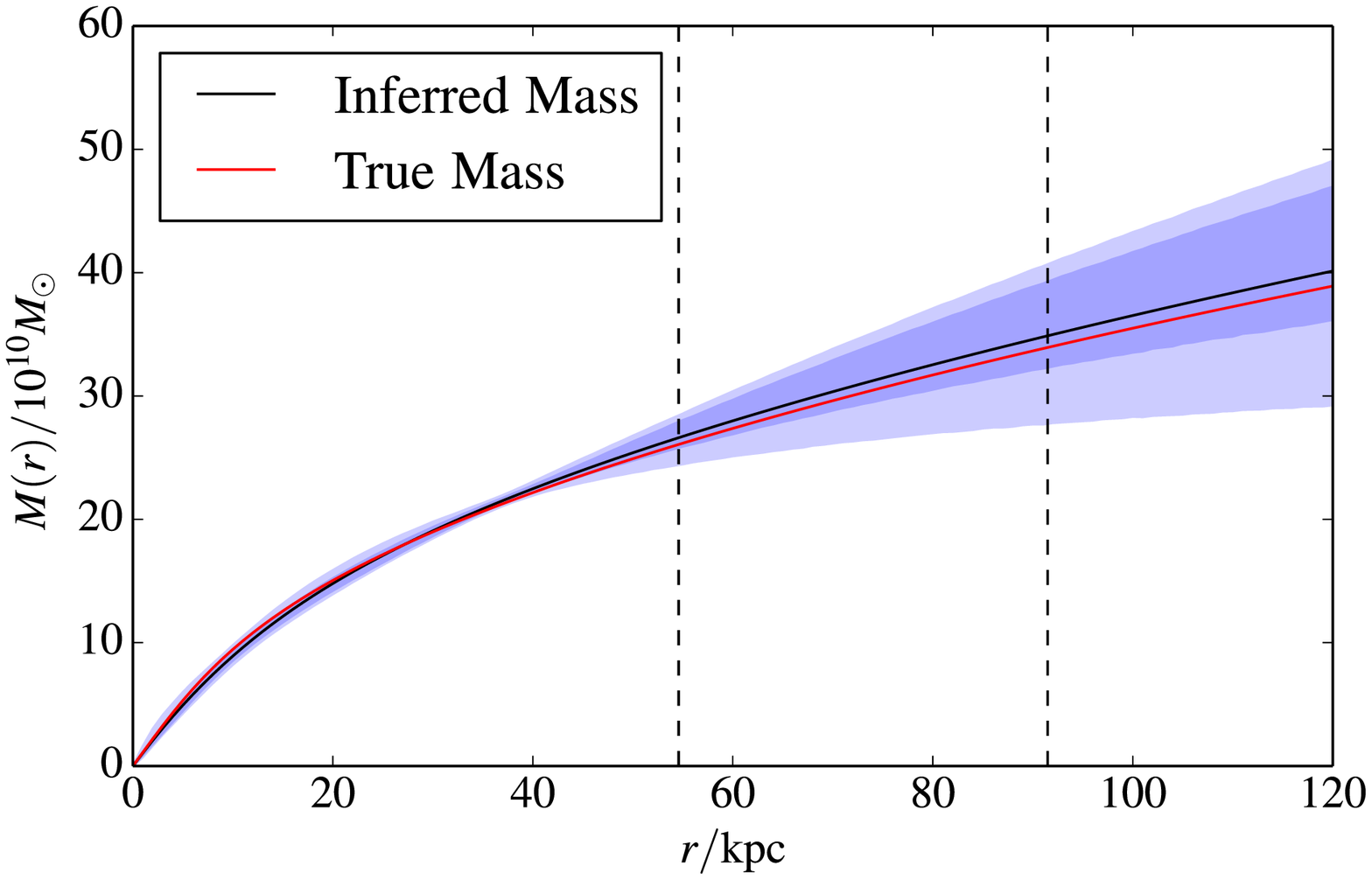}
\includegraphics[width=0.33\textwidth]{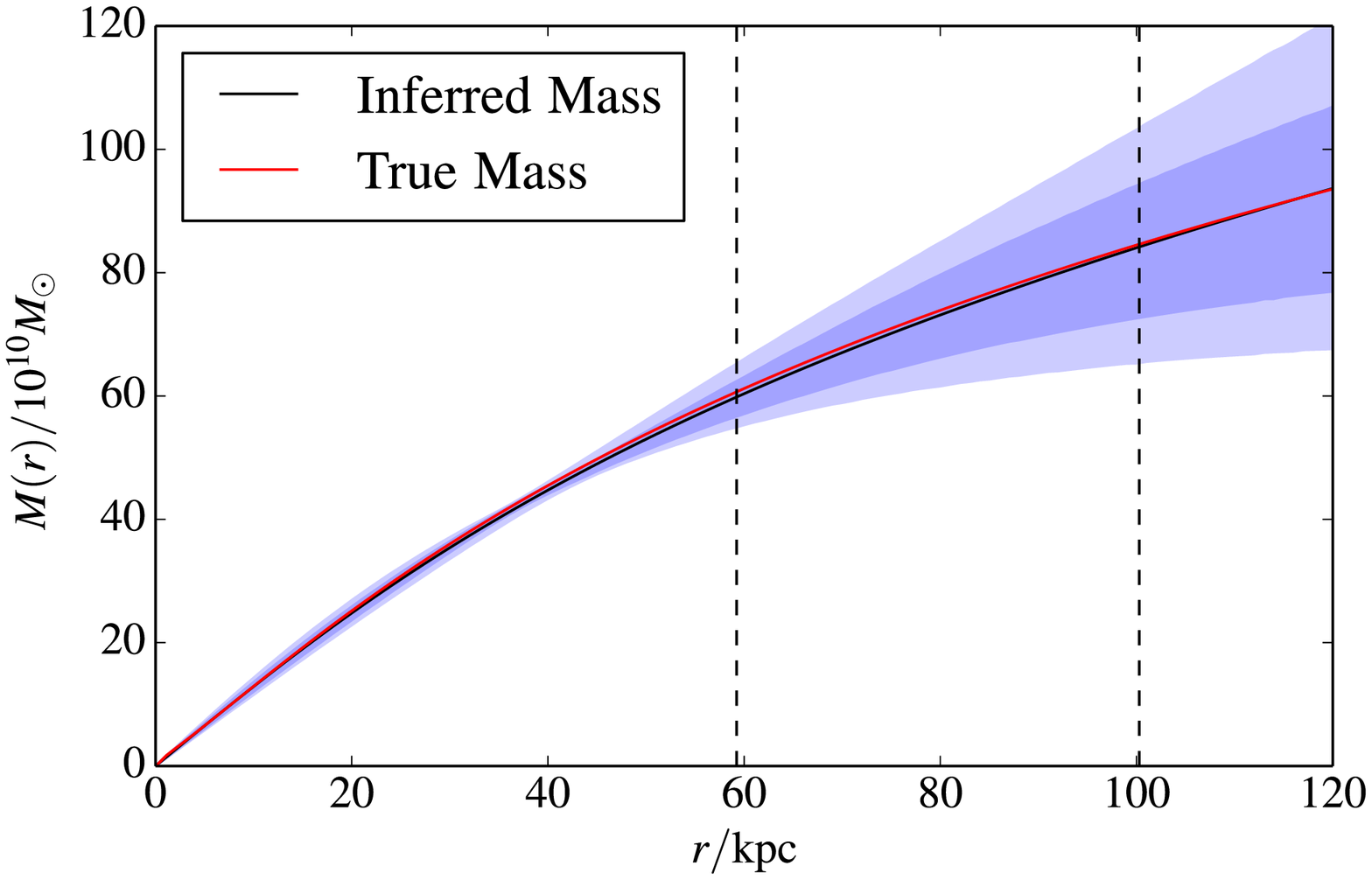}
\includegraphics[width=0.33\textwidth]{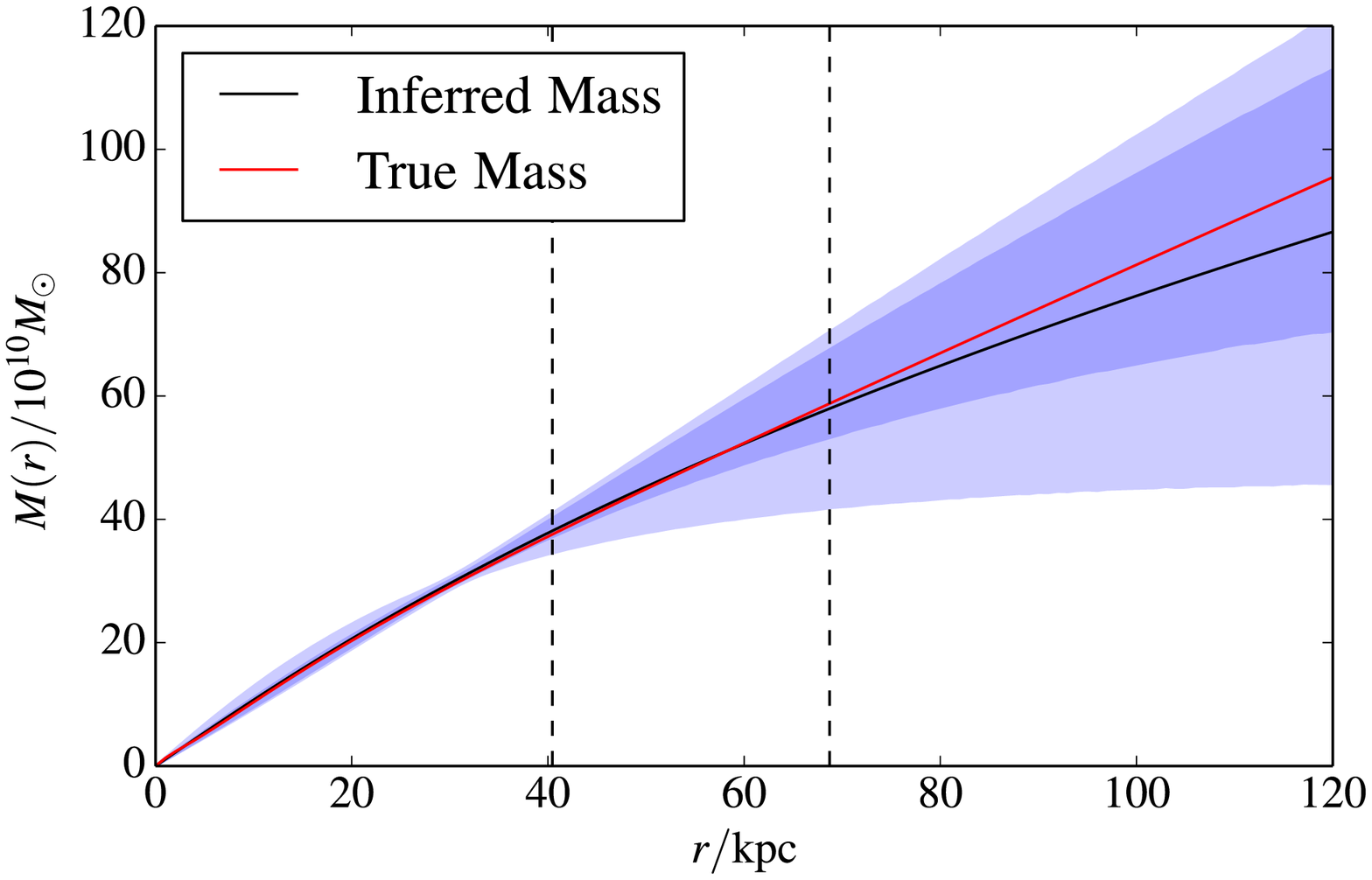}
\caption{\label{fig:mass_recovery} Mass recovery for Sgr disruption in
  three different mock Galaxies. Red curves show the true cumulative
  mass profile, while black lines report the most likely model
  inferences. The dark and light blue regions show the 1 and 2
  $\sigma$ confidence regions of the recovered mass. The vertical
  dashed lines show the radii of the leading and trailing apocentres
  (with $r_{\rm aL} < r_{\rm aT})$.  {\it Left:} The mock Galaxy is
  the spherical TF model. Note the ``pinch'' radius at around 40 kpc
  where the random error is at its minimum. The systematic offset
  between the true and recovered mass is less than 1 standard
  deviation everywhere between 10 and 120 kpc. {\it Middle:}
  Cumulative mass profile for the mock three-component Galaxy with a
  bulge, disc and a spherical dark halo. Results are very similar to
  the test case shown in the left panel, albeit the random error blows
  up slightly faster beyond the pinch radius. {\it Right:} Mass
  distribution in the mock Milky Way with \citet{LM2010} potential. No
  significant systematic offset between data and the model is
  detected. Note, however, that beyond 50 kpc, the error regions are
  the largest of the three cases considered.  This is likely to be due
  to the fact that the stream probes a reduced range of Galactocentric
  radii in this case as shown by the locations of the apocentres.}
\end{figure*}

\subsection{Truncated, Flat Rotation Curve Model}

An N-body simulation of a satellite disruption is carried out in a TF
potential with the following parameters. The amplitude of the circular
velocity curve is $v_0 = 210$ km/s, the scale radius is $r_{\rm s} =
12$ kpc and the outer power law is $\alpha = 0.45$. The apocentres of
the tidal tails were extracted using the method presented in
Section~\ref{sec:extract_apos}, yielding apocentric radii and a
precession angle of $r_{\rm aL} = 54.6 \pm 1.0$~kpc, $r_{\rm aT} =
90.5 \pm 1.2$~kpc and $\Delta \psi = 85^\circ.8 \pm 1.3$.

The resulting mock stream data are then modeled by applying the mLCS
algorithm described in Section 3. In this particular test case, the functional
form of the model potential coincides exactly with that used to produce the
disruption simulation. The left group of panels in
Fig.~\ref{fig:potential_recovery} displays the quality of the inference.
Unsurprisingly, there are strong degeneracies between all three model
parameters, most prominently between $r_{\rm s}$ and $\alpha$. The coupling
between the parameters controlling the steepness of the radial mass density
profile was already apparent in Fig.~\ref{fig:model_deps}. Notwithstanding this
degeneracy, the true parameter values (marked with red stars) all fall within
the uncertainties implied by the posterior distributions.

The accompanying Fig.~\ref{fig:mass_recovery} presents the details of
the mass profile recovery. In particular, the left panel shows the
results for the mock TF Galaxy discussed here. The true cumulative
mass profile is shown in red, while the black curve gives the inferred
mass distribution. The slope of the overall mass (dark matter plus
baryons) profile changes quickly within 10-30 kpc from the Galactic
centre and then stays constant. As is obvious from Fig.~\ref{fig:mass_recovery}, the
mismatch between the true and the inferred profiles is minimal
everywhere within the range considered, i.e. $< 100kpc$. While the
systematic error appears tiny, the random error has a characteristic
``pinch'' radius at around 40 kpc, within which it stays minimal,
i.e. 1\%, and then blows up to 10\% beyond 50 kpc. In other words, the
constraining power of the method lies mostly inside the leading tail
apocentre (the apocentric distance are marked by vertical dashed
lines). Importantly, while the $r_{\rm s}$ and $\alpha$ parameters are
measured with large error-bars due to the above-mentioned degeneracy,
the outer slope of the total radial mass profile is clearly
constrained, as evidenced by the lack of any significant systematic
error between 50 and 100 kpc.

\subsection{Bulge, Disc and Spherical Halo}

While it is reassuring to see that the model performs well in the case
where the functional form of the density and potential distribution is
known, we do not have such a luxury when analyzing the real Milky
Way. Therefore, we proceed to test the algorithm on mock disruption
data produced in Galaxies that differ from the assumed model
distribution.

\begin{figure}
\centering
\includegraphics[width=0.5\textwidth]{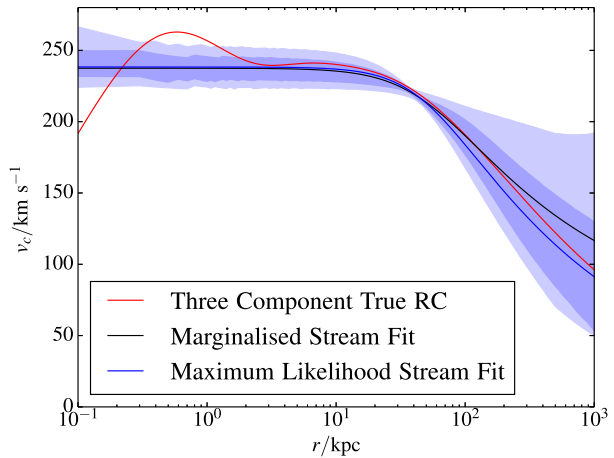}
\caption{\label{fig:vc_test} Rotation curve of the three component
  (bulge, disc and spherical halo) mock Galaxy (red curve). The blue
  line shows the rotation curve based on the maximum likelihood stream
  fit, while the black curve is the marginalized stream fit. It is
  clear that the spherical TF model is a viable choice to describes
  the mass distribution in a more complex potential in the range of
  Galactocentric distances probed by the stream (10~kpc - 100~kpc).}
\end{figure}

Here, in particular, the disruption is produced in the three-component
(bulge, disc and dark halo) mock Galaxy, but the stream fitting is
done using the TF model. The bulge is taken as a \citet{Hernquist1990}
sphere with mass of $M_{\rm b} = 3 \times 10^{10} M_\odot$ and
scale-length $a_{\rm b} = 500~{\rm pc}$. The thick and thin discs are
represented as a single component which follows the
\citet{MiyamotoNagai} model with total mass $M_{\rm d} = 3.3 \times
10^{10} M_\odot$, the scale-length $a_{\rm d} = 4~{\rm kpc}$ and the
scale-hight $b_{\rm d} = 400~{\rm pc}$. Finally, the dark halo is
represented with a spherical NFW model \citep{NFW1996}. The DM halo's
mass is $M_{200} = 1.2 \times 10^{12} M_\odot$ and its concentration
is $c_{200} = 16$. The rotation curve given by this three-component
model is shown in Fig.~\ref{fig:vc_test}. It gives a reasonable
representation of the LSR as measured by \citet{Bovy2012}, with an
amplitude of $\approx 240~{\rm km~s^{-1}}$ around $8~{\rm kpc}$. The
resulting apocentric radii and the precession angle are the following:
$r_{\rm aL} = 57.2 \pm 0.6$~kpc, $r_{\rm aT} = 100.3 \pm 0.3$~kpc and
$\Delta \psi = 110^\circ.6 \pm 0.7$.

As before, the posterior distributions for $r_{\rm s}$, $\alpha$ and
$v_0$ are shown in the middle panels of
Fig.~\ref{fig:potential_recovery}. Unsurprisingly, the signs of the
same degeneracy between $r_{\rm s}$ and $\alpha$ are clearly
visible. It is, however, more difficult to interpret these posterior
distributions in view of the mismatch between the input and the
model. Nonetheless, regardless of the degeneracy and the model
mismatch, the marginalized probability distributions appear to have
well-defined peaks. Most importantly, the cumulative mass profile is
recovered with inspiring fidelity as evidenced by the middle panel of
Fig.~\ref{fig:mass_recovery}. As in the previous test case, the
posterior distribution of the cumulative mass profile displays a
``pinch'' at around 40 kpc. Note that the two mock streams actually
have similar apocentric distances. The behaviour of the random error
beyond 50 kpc is slightly different: it grows to somewhat larger value
13\%, which we consider to by symptomatic of the mismatch between the
``true'' and the assumed Galactic mass distributions. Interestingly,
the rotation curve inferred from the stream apocentric data looks a
very good match to the input one as seen in Fig.~\ref{fig:vc_test}.

\subsection{Model of \citet{LM2010}}

It is clear from the two test cases discussed above that the algorithm
can be used to infer the cumulative mass distributions in both
spherical and flattened Galaxies. However, how does it cope with a
yet more complex potential?

To this end, we produce an N-body simulation of a satellite disruption
mirroring the setup of \citet{LM2010}. The result of their modeling of
the Sgr stream data available at the time is intriguing. Together with
the usual components for the bulge/bar and the discs, their model
favors a DM halo that appears almost perfectly oblate, but with the
minor axis stuck in the Galactic plane. The combined gravity of the
spherical bulge, highly flattened disc and the ``hockey puck'' halo
results in a complex overall potential with the shape evolving between
5 and 60 kpc.

The outcome of our N-body simulations looks identical to the snapshots
published by \citet{LM2010}. In particular, for the Sgr stream we find
the apocentric radii of following: $r_{\rm aL} = 47.5 \pm 1.3$~kpc,
$r_{\rm aT} = 68.7 \pm 1.4$~kpc and the precession angle is
$\Delta \psi = 114^\circ.1 \pm 3.0$. Note that these values are
markedly different from the early two cases and from those measured as
pointed out by \citet{Belokurov2014}. The result of applying the model
of a stream forming in the TF potential to these mock data are show in
the right panel of Fig.~\ref{fig:potential_recovery}. The model
parameters appear to be even less constrained in this case, although,
the marginalized posterior probability distributions for $r_{\rm s}$ and
$v_0$ do show very clear peaks.

The most important conclusion, however, can be gleaned from the right
panel of Fig.~\ref{fig:mass_recovery} which proves that the Galactic
total mass and the radial matter distribution can be measured even in
such a complicated potential. It is true that both systematic and
random error (which is at 20\% level at 100 kpc) in this case are
larger than the previous two. The ``pinch'' radius has moved in and is
now at around 30 kpc, but this is simply the consequence of the
smaller apocentric radii of the mock stream. The mismatch between the
true and the inferred mass starts to grow quickly beyond 60 kpc, but
stays comfortably within one standard deviation.

\section{Application to the Milky Way Galaxy}\label{sec:MW_application}

\begin{figure}
\centering
\includegraphics[width=0.5\textwidth]{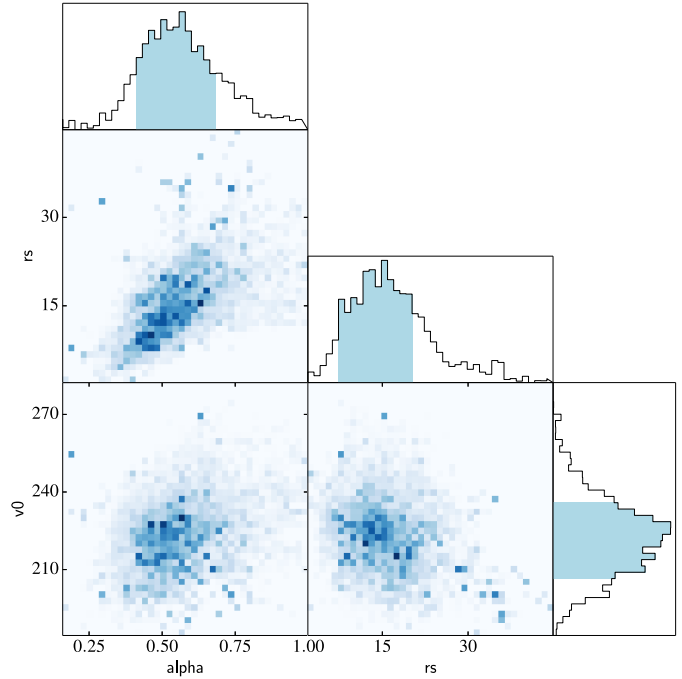}
\caption{\label{fig:MW_potential} Posterior probability distributions
  for $r_s-\alpha$ parameters in the TF model as inferred from the Sgr
  stream precession. Note -- as seen previously in the tests with mock
  Galaxies -- that there is a degeneracy between the two shape
  parameters of the model. However, the degeneracy appears more
  bloated, most likely due to the uncertainty in the determination of
  the 6D position of the Sgr dwarf. Histograms show the marginalized
  posterior for each of the parameters.}
\end{figure}

With confidence in our model's ability to perform unbiased inference,
we now turn to the task of fitting the data for the apocentric
distances and precession angle of the Sgr stream.  The priors used on
each of the model parameters are tabulated in Table~2.  We use
completely uninformed priors on each of the parameters of the Galaxy
model, allowing this method to give us an independent measure of the
Milky Way's potential.

\begin{table}
\centering
\begin{minipage}{80mm}
\caption{The priors on each of the model parameters when fitting to
  the Sgr stream data in section~\ref{sec:MW_application}.  When a
  uniformly distributed prior is used, the range is indicated with
  square brackets. When a normally distributed prior is used, the mean
  and standard deviation is indicated as a pair $(\mu, \sigma)$ in
  round brackets.}  \centering
\begin{tabular}{ccc}
\hline
{\bf Parameter} & {\bf Distribution} & {\bf Prior}\\
\hline
\multicolumn{3}{l}{Potential Parameters} \\
\hline
$v_0$ & Uniform & $[40, 400]~{\rm km~s^{-1}}$\\
$r_s$ & Uniform & $[1,100]~{\rm kpc}$\\
$\alpha$ & Uniform & [0,1] \\
\hline
\multicolumn{3}{l}{Progenitor Properties} \\
\hline
$m_{\rm sat}$ & Uniform & $[0.1, 1] \times 10^9 M_\odot $ \\
$a_{\rm sat}$ & Uniform & $[0.1, 1]~{\rm kpc}$\\
$\sigma$ & Uniform & $[0.5, 10.0]~{\rm km~s^{-1}}$\\
\hline
\multicolumn{3}{l}{Progenitor's Orbit} \\
\hline
$l$ & Fixed & $5^\circ.5689$\\
$b$ & Fixed & $-14^\circ.1669$\\
$d_{\rm helio}$ & Uniform & $[22.0, 28.4]~{\rm kpc}$\\
$v_{\rm los}$ & Normal & $(153, 2)~{\rm km~s^{-1}}$\\
$\mu_l \cos b$ & Normal & $(1.97, 0.3)~{\rm mas~yr^{-1}}$\\
$\mu_b$ & Normal & $(-2.44,0.3)~{\rm mas~yr^{-1}}$ \\
$t_{\rm back}$ & Uniform & [0, 10] Gyr \\
\hline
\multicolumn{2}{l}{LSR Properties} \\
\hline
$R_0$ & Fixed & $8.0~{\rm kpc}$ \\
$v_c(R_0)$ & Fixed & $237.0~{\rm km~s^{-1}}$\\
$U_\odot$ & Fixed & $11.1~{\rm km~s^{-1}}$\\
$V_\odot$ & Fixed & $12.24~{\rm km~s^{-1}}$\\
$W_\odot$ & Fixed & $7.25~{\rm km~s^{-1}}$\\
\hline
\end{tabular}
\end{minipage}
\label{PriorTable}
\end{table}

\begin{table}
\centering
\begin{minipage}{80mm}
\centering
\caption{The enclosed mass of the Milky Way as inferred from the
  stream precession modeling.  We provide estimates at 50, 100, 150
  and 200~kpc. Along with 68\% and 95\% confidence intervals.}
\begin{tabular}{cccc}
\hline
$r/{\rm kpc}$ & $M(r) / 10^{11} M_\odot$ & $1\sigma / 10^{11} M_\odot$ & $2\sigma / 10^{11} M_\odot$ \\
\hline
50  & 2.9 & 0.4 & 0.9\\
100 & 4.1 & 0.7 & 1.6\\
150 & 4.9 & 1.0 & 2.4\\
200 & 5.6 & 1.2 & 3.0\\
\hline
\end{tabular}
\end{minipage}
\label{MassTable}
\end{table}

\begin{figure*}
\centering
\includegraphics[width=0.49\textwidth]{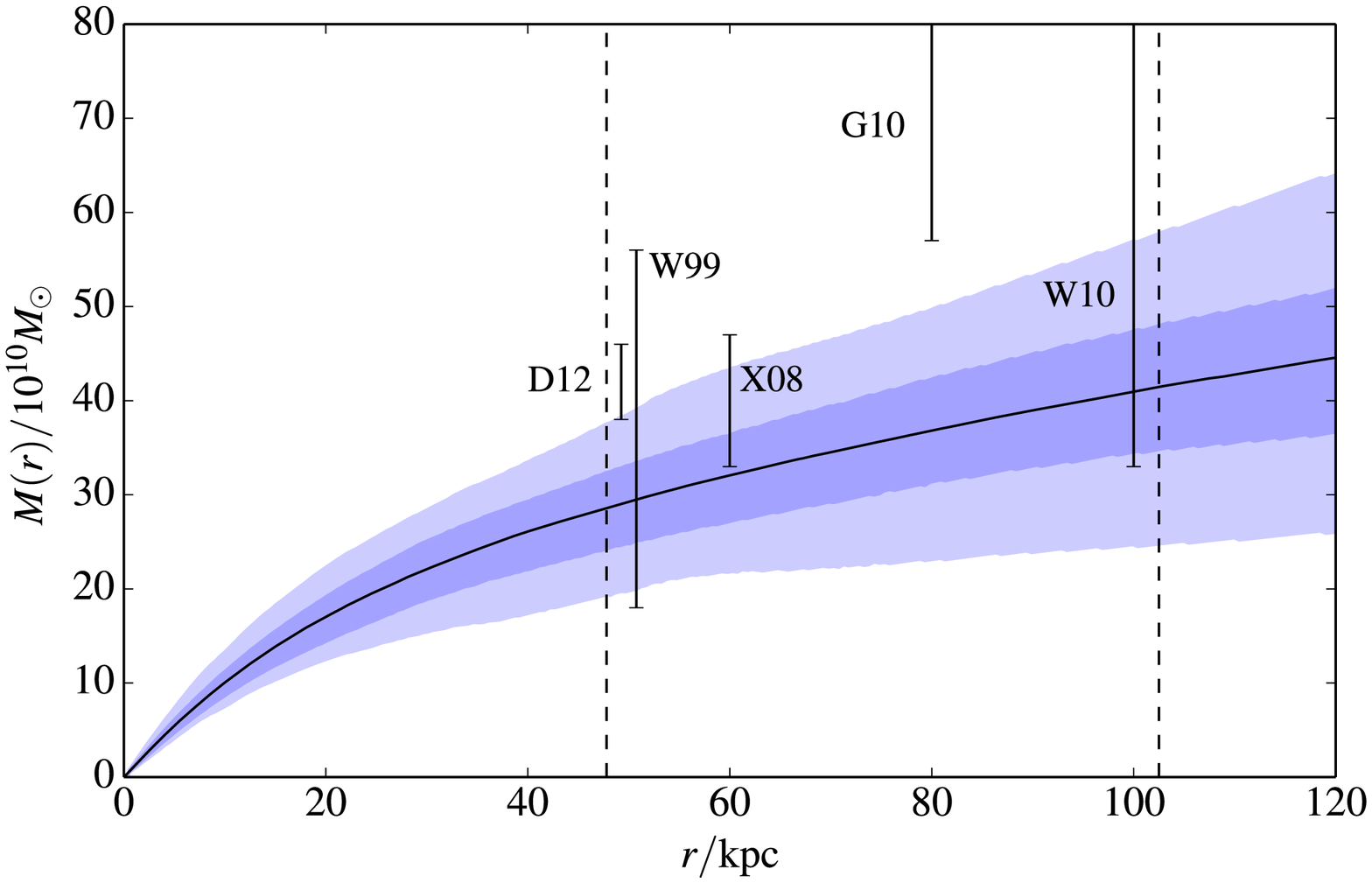}
\includegraphics[width=0.49\textwidth]{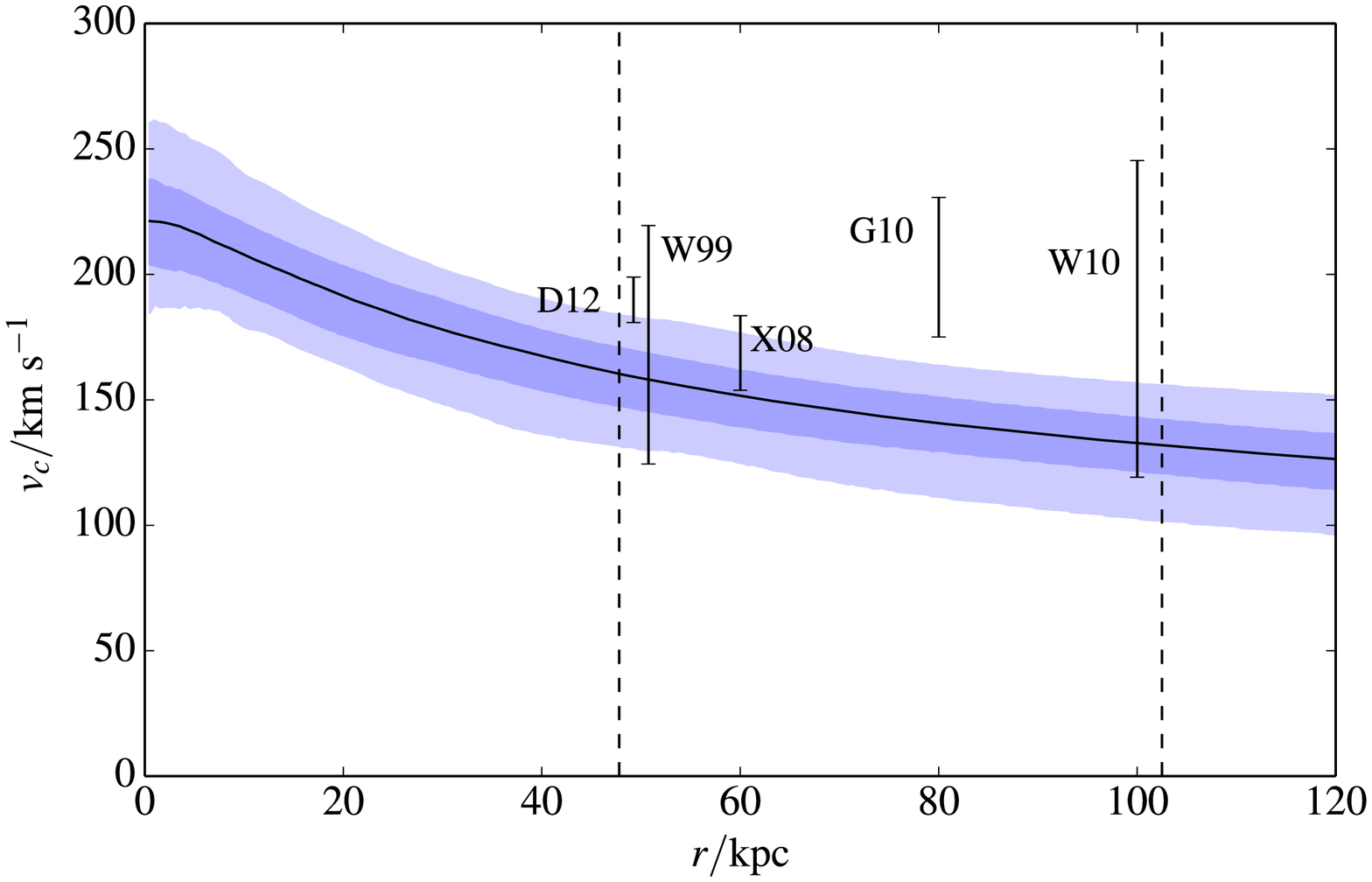}
\caption{\label{fig:MW_mass} {\it Left:} Cumulative mass profile of
  the Milky Way as inferred by the model of the Sgr stream
  precession. The solid black line shows the most likely mass, the
  light and dark blue shaded regions show the 68\% and the 95\%
  confidence regions respectively.  The black points are previous
  determinations from the literature and are labelled as follows: D12
  -- \citet{Deason2012BD}; G10 -- \citet{Gnedin2010}; X08 --
  \citet{Xue2008}; W99 -- \citet{Wilkinson1999}; W10 --
  \citet{Watkins2010}. The D12 and W99 points at 50~kpc have been
  offset from each other for clarity. Note that the error bars for G10
  and W10 extend up to $9.9\times10^{11} M_\odot$ and $13.8 \times
  10^{11} M_{\odot}$ respectively. Our mass determination agrees well
  with most of the literature values, but systematically favors the
  lower margin of the reported confidence intervals {\it Right:} Milky
  Way circular velocity curve as inferred by the model. The solid
  black line shows the most likely $V_{\rm c}$, the light and dark
  blue shaded regions show the 68\% and the 95\% confidence regions
  respectively.}
\end{figure*}

The inference on the potential parameters is presented in
Fig.~\ref{fig:MW_potential}. Here, the same general structure as with
the test cases can be seen, with the obvious degeneracy between
$r_{\rm s}$ and $\alpha$. In this case, the narrow banana-shaped
clouds are significantly broadened due to the noise in the
determination of the current phase-space position of the progenitor.
The inferred mass profile and rotation curve of the Milky Way are
shown in Fig.~\ref{fig:MW_mass}, alongside a selection of recent mass
determinations from the literature. According to our model, the
cumulative mass profile flattens out at around 40 kpc, and from there
onwards remains rather shallow. If the matter density profile actually
does not change slope beyond 80-100 kpc, the total Milky Way mass
ought to be rather low, i.e. $0.5-0.8 \times 10^{12}
M_{\odot}$. Indeed, quantitatively, our measurement appears to be at
the lower bound of most previous estimates. We tabulate our mass
estimates for a range of radii in Table 3.

At $50~{\rm kpc}$ we are in good agreement with the mass determined by
\citet{Wilkinson1999} from satellite and globular cluster
motions. There is, however some tension with the measurement from BHB
tracers of \citet{Deason2012BD}, where agreement is only at the
2$\sigma$ level. We are in excellent agreement with the mass
measurement out to 60~kpc of \citet{Xue2008}. There is also tension
with the measurement of \citet{Gnedin2010} of the mass out to
80~kpc. This discrepancy is easy to explain. The analysis is based on
the sample of hyper-velocity stars published by
\citet{Brown2010}. However, a staggering 50\% of these are actually
projected to lie in the Sgr stream. The already crippling effects of
soaring Sgr stream contamination are possibly exacerbated by
additional contamination from Blue Stragglers. A more recent study of
the kinematics of the outer stellar halo tracers by \citet{ColdHalo}
finds evidence of a dramatic drop in the velocity dispersion beyond 50
kpc. Superficially, their measurement of the line-of-sight velocity
dispersion beyond 100 kpc, $\sigma_r\sim 60$ kms$^{-1}$ is consistent
with our measurement of the circular velocity at similar distances,
$V_c\sim 120$ kms$^{-1}$.

Finally, we compare our results to those of \citet{Watkins2010}. They
performed inference based upon the satellite population of the Milky
Way, again assuming smooth density laws. Their results are strongly
dependent on the (unknown) velocity anisotropy of their population,
giving a wide range of possible masses. They find that the plausible
range of masses out to 100~kpc lies between 3.3 and 13.8$\times
10^{11} M_\odot$.  The results of this work favour the lower end of
their range.

\section{Discussion and Conclusions}\label{sec:discussion_conclusion}

We have developed a rapid algorithm capable to generate
realistically looking streams produced in disruption of progenitors of
arbitrary mass in arbitrary potentials. In our model, stars are
released with a Gaussian velocity distribution at the inner and outer
Lagrange points of the disrupting progenitor. An essential ingredient
is the inclusion of the effect of the progenitor's gravity on stream
particles around the time of un-binding.  In essence, the progenitor's
gravity naturally corrects our crude guess for the velocity
distribution of stripped particles to a more realistic one.

This model provides a natural explanation for the differences in the
apocentric distances and the precession angle found for the
Sagittarius (Sgr) stream in \citet{Belokurov2014}.  The behaviour of
the stream properties is controlled mostly by the host potential in
which the satellite is disrupting, though there is a small dependence
upon the internal properties of the progenitor. Thus, the observations
are explicable without the need to invoke dynamical
friction~\citep[c.f.,][]{Chakrabarti2014}.

Exploiting the sensitivity of the precession measurement to the
underlying potential, we tested the ability of our model to infer the
potential from mock observations from N-body simulations. We find the
model performs remarkably well and produces a nearly unbiased estimate
of the enclosed mass.  When applied to observations of the precession
of the Sgr Stream, we obtain a new measurement of the mass profile of
the Milky Way out to distances of $\sim 100~{\rm kpc}$. We find the
masses within 50~kpc and 100~kpc to be $M(50~{\rm kpc}) = 2.9 \pm
0.5 \times 10^{11} M_\odot$ and $M(100~{\rm kpc}) = 4.0 \pm 0.7 \times
10^{11} M_\odot$ respectively. We emphasise that this is an entirely
independent method to any previous determinations.

What are the limitations of our method? Ideally, when modelling the
satellite disruption, one should aim at reproducing the properties of
both the stream as well as the remnant. By design, the mLCS algorithm
presented in this paper has nothing to say about the final state of
the stream progenitor. Additionally, while we have established that
the centroids of streams generated by Modified Lagrange Cloud
Stripping matches well the centroids of streams produced in N-body
simulation, there is no evidence that our model gives realistic
density distribution along or the tidal tails. This is because in the
current implementation, the stripping rate is independent of
time. However, as the N-body simulations show the stellar flux out of
the satellite is a strong function of time. Moreover, if the
progenitor contains both stars and dark matter, the stellar stripping
rate is highly suppressed in the initial throes of accretion, and
sharply increases once most of the DM has been removed \citep[see
  e.g.][]{taletwicetold}. Finally, if the Galactic dark matter halo is
strongly triaxial, it might be possible to bias our stream-based
inference of the total mass. We have explored how the assumption of
sphericity of the overall potential influences the mass
recovery. However, none of our mock Galaxies are truly triaxial, for
example, the dark halo in the model of \citet{LM2010} is nominally
triaxial, but in reality it is just a oblate ellipsoid standing on a
side.

Our results represent another piece in the growing body of
evidence \citep{Battaglia2005, Bovy2012, Deason2012BD, Rashkov2013}
that suggests the Milky Way galaxy is less massive than previously
assumed. The total mass of the Milky Way inferred from the kinematics
of satellites depends critically on whether the distant and fast
moving dwarf galaxy Leo I is included or not ~\citep[see
e.g.,][]{Wilkinson1999,Watkins2010}.  If Leo I is unbound, then
$M(200~{\rm kpc}) \lesssim 1 \times 10^{11} M_\odot$, whilst if Leo I
is bound, then $M(200~{\rm kpc}) \lesssim 2 \times 10^{12}
M_\odot$.  \citet{Boylan2013} argue that Leo I is most likely bound,
as 99.9 per cent of sub-haloes in their simulations are bound to the
host.  Nonetheless, processes to create fast-moving satellites are
known. For example, infall of a satellite pair onto a host may cause
the heavier satellite to remain bound whilst the lighter satellite is
ejected. \citet{Sales07} find that that as many as a third of all
satellites in their suite of simulations lie on such orbits. Although
such extreme satellites may still be bound, the important point is
that they do not constitute part of the virialized population. Mass
estimates of the Milky Way using satellites depend on applying the
virial theorem or the Jeans equations to the satellite populations
assuming time-independence. Extreme satellites produced by three-body
interactions should not be included in the sample.

Despite the advocacy of \citet{Boylan2013}, there is another strong
reason to exclude Leo I from the sample of bound satellites. Then,
three independent methods of estimating the mass of the Milky Way --
namely from the kinematics of distant halo
stars~\citep{Deason2012BD,Rashkov2013}, the kinematics of the
satellite galaxies~\citep{Watkins2010} and the modelling of the Sgr
stream (this paper) -- are all in good agreement. They all suggest
that the total mass of the Milky Way is $\lesssim 1\times 10^{12}
M_\odot$. It is important to realise that high mass estimates for the
Milky Way galaxy essentially depend on a single datapoint, namely the
inclusion of Leo I in the sample of satellites galaxies modelled as a
virialized population.

This has substantial implications for one of the alleged problems of
$\Lambda$CDM, the \lq\lq Too Big to Fail
Problem". \citet{BoylanKolchin2011} provide a lucid articulation of
the problem.  For simulated Milky Way analogues of mass $\sim 2 \times
10^{12} M_\odot$, the most massive subhaloes are too dense to
correspond to any of the known satellite galaxies of the Milky
Way. They typically have peak circular velocities of 30 kms$^{-1}$,
which is too large to plausibly correspond to the most luminous dwarf
spheroidal satellites of the Milky Way. Baryonic feedback does not
appear to solve the problem entirely~\citep{Ga13}, so many researchers
have interpreted this as evidence for changing the nature of the dark
matter particle to warm~\citep{L012} or self-interacting~\citep{Vo12}
or asymmetric~\citep{Zu14}. However, by far the simplest and likeliest
way to resolve the ``Too Big to Fail Problem'' is to reduce the mass
of the Milky Way Galaxy to $\sim 1 \times 10^{12} M_\odot$.

The work in this paper now provides a completely new and independent
line of argument supporting a much leaner Milky Way Galaxy. Certainly
out to $\sim 100$ kpc, the Sgr stream provides a particularly clean
tool for mass estimation of the Milky Way. It is preferable to
analyses of the satellite galaxies, as it circumvents the problem of
sample contamination by unbound or unvirialized objects.  It is
preferable to analyses of the kinematics of halo stars as there is no
mass-anisotropy degeneracy to frustrate mass determinations. We
anticipate that -- with increasing quality and quantity of data over
the next few years -- it will become the gold standard for mass
measurements of the Milky Way.

\section*{Acknowledgements}
The authors wish to thank the anonymous referee for a helpful and constructive
report.  The authors would also like to thank the members of the Cambridge
``Streams'' discussion club for their invaluable contribution. SG thanks the
Science and Technology Facilities Council (STFC) for the award of
a studentship. VB acknowledges financial support from the Royal Society. The
research leading to these results has received funding from the European
Research Council under the European Union's Seventh Framework Programme
(FP/2007-2013) / ERC Grant Agreement n. 308024.

\label{lastpage}

\bibliographystyle{mn2e}
\bibliography{refs}

\end{document}